\documentstyle[12pt,axodraw,epsfig]{article}
        
\def\beq{\begin{equation}}
\def\eeq#1{\label{#1}\end{equation}}
\def\eeqn{\end{equation}}
\def\beqa{\begin{eqnarray}}
\def\eeqa#1{\label{#1}\end{eqnarray}}
\def\eeqan{\end{eqnarray}}
\def\CR{\nonumber \\ }
\def\leqn#1{(\ref{#1})}

\def\({\left(}
\def\){\right)}
\def\B{B_H}
\def\plus{+}
\def\minus{-}

\def\to{\rightarrow}

\def\cw#1{\cos^{#1}\theta_W}

\newcommand{\mN}{m_n}

\newcommand{\text}[1]{\rm #1}

\newcommand{\sla}[1]{\not{\! #1}}

\newcommand{\eqref}[1]{Eq.~(\ref{#1})}

\def\stacksymbols #1#2#3#4{\def\theguybelow{#2}
    \def\vp{\lower#3pt}
    \def\sp{\baselineskip0pt\lineskip#4pt}
    \mathrel{\mathpalette\intermediary#1}}

\def\intermediary#1#2{\vp\vbox{\sp
     \everycr={}\tabskip0pt
     \halign{$\mathsurround0pt#1\hfil##\hfil$\crcr#2\crcr
              \theguybelow\crcr}}}

\def\gapproxeq{\stacksymbols{>}{\sim}{2.5}{.2}}
\def\lapproxeq{\stacksymbols{<}{\sim}{2.5}{.2}}

\begin{document}

\begin{titlepage}
\begin{flushright}
{\tt hep-ph/yymmnnn} \\
\end{flushright}

\vskip.5cm
\begin{center}
{\huge \bf Little Higgs Dark Matter} \vskip.2cm
\end{center}
\vskip1cm

\begin{center}
{\bf Andreas Birkedal$^{1}$, Andrew Noble$^{2}$, Maxim Perelstein$^2$, \\
and Andrew Spray$^{2}$} \\
\end{center}
\vskip 8pt

\begin{center}
{\it $^1$ SCIPP, University of California, Santa Cruz, CA~95064\\
$^2$ Cornell Institute for High-Energy Phenomenology, Cornell University, Ithaca, NY~14853}
\end{center}

\vglue 0.3truecm

\begin{abstract}
\vskip 3pt \noindent
The introduction of T parity dramatically improves the consistency of Little
Higgs models with precision electroweak data, and renders the lightest T-odd
particle (LTP) stable. In the Littlest Higgs model with T parity, the LTP is 
typically the T-odd heavy photon, which is weakly interacting and can play 
the role of dark matter. We analyze the relic abundance of the heavy photon,
including its coannihilations with other T-odd particles, and map out the
regions of the parameter space where it can account for the observed dark 
matter. We evaluate the prospects for direct and indirect discovery of the
heavy photon dark matter. The direct detection rates are quite low and 
a substantial improvement in experimental sensitivity would be required 
for observation. A substantial flux of energetic gamma rays is produced in 
the annihilation of the heavy photons in the galactic halo. This flux 
can be observed by the GLAST telescope, and, if the distribution of dark  
matter in the halo is favorable, by ground-based telescope arrays 
such as VERITAS and HESS.
\end{abstract}

\end{titlepage}





\section{Introduction} 

It has now been firmly established that about 25\%
of the energy density in the universe exists in the form of nonrelativistic, 
non-baryonic, non-luminous matter, so called ``dark matter''~\cite{DMreview}. 
The microscopic composition of dark matter remains a mystery, but it is clear 
that it cannot consist of any elementary particles that have been directly 
observed in the laboratory so far\footnote{In principle, it remains possible 
that dark matter consists of microscopic black holes made out of ordinary 
particles. However, we do not know of a compelling cosmological scenario in 
which this possibility is realized.}. Many theories which extend the standard 
model (SM) 
of electroweak interactions contain new particles with the right properties 
to play the role of dark matter; perhaps the best known example is the 
lightest neutralino of supersymmetric (SUSY) models. 

Recently, a new class of theories extending the SM at the TeV scale, 
``Little Higgs'' (LH) models, has been proposed~\cite{littlest} (for reviews, see~\cite{LHrev1,LHrev2}). The LH 
models contain a light (possibly composite) Higgs boson, as well as additional 
gauge bosons, fermions, and scalar particles at the TeV scale. The Higgs is a 
pseudo-Nambu-Goldstone boson, corresponding to a global symmetry spontaneously 
broken at a scale $f\sim 1$ TeV. The global symmetry is also broken explicitly 
by the gauge and Yukawa couplings of the Higgs. As a result of this breaking,
the Higgs acquires a potential; however, the leading (one-loop, quadratically 
divergent) contribution to this potential vanishes due to the special
``collective'' nature of the explicit global symmetry breaking, and the
lightness of the Higgs can be achieved without fine-tuning. The dynamics of 
the Higgs and other degrees of freedom relevant at the TeV scale is described 
by a non-linear sigma model (nlsm), valid up to the cutoff scale 
$\Lambda\sim 4\pi f\sim 10$ TeV. In particular, the Higgs
mass term is dominated by a one-loop, logarithmically enhanced contribution
from the top sector, which can be computed within the nlsm 
and shown to have the correct sign to trigger electroweak symmetry 
breaking, providing a simple and attractive explanation of this 
phenomenon. Above the cutoff scale, the 
model needs to be embedded in a more fundamental theory; however, for many
phenomenological applications, including the analysis of this paper, the
details of that theory are not relevant and the nlsm description suffices.

The Littlest Higgs model~\cite{littlest} is simple and economical, and it has
been the focus of most phenomenological analyses to date~\cite{pheno}. 
Unfortunately, the 
model suffers from severe constraints from precision electroweak fits, due
to the large corrections to low-energy observables from the tree-level
exchanges of the non-SM TeV-scale gauge bosons and the small but non-vanishing 
weak-triplet Higgs vacuum expectation value (vev)~\cite{PEW}.  To 
alleviate this difficulty, the symmetry of the theory can be enhanced to 
include a $Z_2$ discrete symmetry, named ``T parity''~\cite{LHT}. In the
Littlest Higgs model with T parity (LHT)~\cite{Low}, the non-SM gauge 
bosons and the triplet Higgs are T-odd, forbidding all tree-level corrections 
to precision electroweak observables\footnote{In the version of the 
model considered here, there is one non-SM T-even state, the ``heavy top'' 
$T_+$. However it only contributes at tree level to observables involving the 
weak interactions of the top 
quark, which are at present unconstrained.}. Loop corrections to precision 
electroweak observables in the LHT model were considered in~\cite{HMNP}, and 
the model was shown to give acceptable electroweak fits in large regions of
parameter space compatible with naturalness. 

An interesting side effect of T parity is that the lightest T-odd particle 
(LTP) is guaranteed to be stable. Analyzing the spectrum of the model, 
Hubisz and Meade~\cite{HM} have argued that the LTP is likely to be the
electricaly neutral, weakly interacting ``heavy photon'' (or, 
more precisely, the T-odd partner of the hypercharge gauge boson) $B_H$. This 
particle is an attractive dark matter candidate, and initial 
calculations~\cite{HM} showed that its relic abundance is within the observed 
range for reasonable choices of model parameters.\footnote{While the LHT 
dark matter candidate is a spin-1 heavy photon, this is not an unambiguous 
prediction of Little Higgs models. For example, the ``Simplest Little Higgs'' 
models~\cite{Simplest} supplemented by T-parity may contain a stable heavy 
neutrino which can play the role of dark matter~\cite{Martin}, while
closely related ``theory space'' models can give rise to a scalar WIMP dark 
matter candidate~\cite{AJ}.}
In this paper, we will present a somewhat more detailed relic 
density calculation, including the possibility of coannihilations between 
the $B_H$ and other T-odd particles. We will then discuss the prospects for 
direct and indirect detection of the heavy photon dark matter.

\section{The Model} 
\label{model}

Our analysis will be performed within the framework of the
Littlest Higgs model with T parity, which has recently been studied in 
Refs.~\cite{HMNP,HM}. Let us briefly sketch the salient features of the 
model relevant here; for more details, see~\cite{HMNP,HM} or the  review article~\cite{LHrev2}.

The model is based on an $SU(5)/SO(5)$ global symmetry breaking pattern; the 
Higgs doublet of the SM is identified with a subset of the Goldstone boson 
fields associated with this breaking. The symmetry breaking occurs at a scale 
$f\sim 1$ TeV. An $[SU(2)\times U(1)]^2$ 
subgroup of the $SU(5)$ is gauged; this is broken at the scale $f$ down to 
the diagonal subgroup, $SU(2)_L\times U(1)_Y$, identified with the SM 
electroweak gauge group. The extended gauge structure results in four 
additional gauge bosons at the TeV scale, $W^\pm_H$, $W^3_H$ and 
$B_H$.\footnote{ 
The $W_H^3$ and $B_H$ fields mix to form the two neutral mass eigenstates;
however, the mixing angle is of order $v/f$ and can typically be neglected.}  

T parity is an automorphism which exchanges the 
$\left[SU(2) \times U(1) \right]_1$ and $\left[SU(2) \times U(1) \right]_2$ 
gauge fields; under this transformation, the TeV-scale gauge bosons are odd, 
whereas the SM gauge bosons are even. The odd gauge bosons have masses
\beq
M(W_H^a)\,\approx\, gf,~~~M\equiv M(B_H)\,\approx\, 
\frac{g^\prime f}{\sqrt{5}}\approx 0.16 f,
\eeq{mass}
where $g$ and $g^\prime$ are the SM $SU(2)_L$ and $U(1)_Y$ gauge couplings,
and the normalization of $f$ is the same as in Ref.~\cite{HMNP}. 
(Electroweak symmetry breaking at the scale $v\ll f$ induces corrections 
to these formulas of order $v^2/f^2$.) The ``heavy photon'' $B_H$ is the 
lightest new gauge boson, and in fact is quite light compared to $f$. Since
the masses of the other T-odd particles are generically of order $f$, we
will assume that the $B_H$ is the lightest T-odd particle (LTP), and it will 
play the role of dark matter candidate. The only direct coupling of the 
heavy photon to the SM sector is via the Higgs, resulting in weak-strength 
cross sections for $\B$ scattering into SM states. The heavy photon then
provides yet another explicit example of a weakly interacting massive 
particle (WIMP) dark matter candidate, and it is not surprising that we will
find reasonable regions of parameter space where it can account for all of
the observed dark matter. For later convenience, we denote
the mass of this particle by $M$. The range of the allowed values for this parameter is determined by the precision electroweak constraints, which put a lower bound on $f$, typically of about 600 GeV~\cite{HMNP}. While there is no firm upper bound on $f$, we will assume $f\lapproxeq 2$ TeV to avoid reintroducing fine tuning in the Higgs sector. Using Eq.~\leqn{mass}, this corresponds to the WIMP masses in the range
\beq
100~{\rm GeV} \lapproxeq M \lapproxeq 300~{\rm GeV}.
\eeq{WIMPmassrange}
In the scalar sector, the model 
contains an additional T-odd weak-triplet field $\phi$, which has a mass
of order $f$ and no vacuum expectation value. In the fermion sector, each SM 
doublet ($Q_i^a$ and $L_i$, where $a=1\ldots 3$ is a color index and 
$i=1\ldots 3$ is a generation index), acquires a T-odd partner, 
$\tilde{Q}_i^a$ and $\tilde{L}_i$. The masses of these particles are also 
free parameters,\footnote{If the flavor structure 
of the T-odd quark mass matrix is generic, with order-one flavor mixing 
angles, the masses of the T-odd quarks need to be degenerate at the few per
cent level~\cite{LHTflavor}.} with the natural scale set by $f$. To   
avoid proliferation of parameters, we will assume a universal T-odd fermion 
mass $\tilde{M}$ for both lepton and quark partners; we
will require $\tilde{M}>M$ to avoid charged or colored LTPs, and 
assume $\tilde{M}\gapproxeq 300$ GeV, since otherwise the colored T-odd 
particles would 
have been detected in the squark searches at the Tevatron. In addition,
non-observation of four-fermion operator corrections to SM processes
such as $e^+e^-\to q\bar{q}$ places an {\it upper} bound on the T-odd  
fermion masses~\cite{HMNP}: 
\beq
\tilde{M}_{\rm TeV} < 4.8 \,f^2_{\rm TeV}\,,
\eeq{upper}
where $\tilde{M}$ and $f$ are expressed in units of TeV. To cancel the 
one-loop quadratic divergence in the Higgs mass due to top loops, two 
additional new fermions are required in the top sector, the T-even $T_+$
and the T-odd $T_-$.\footnote{In Ref.~\cite{noplus}, a variation of the model
has been constructed where a single T-odd top partner is sufficient to 
cancel the divergences. Since the top sector will only play a minor role in 
the analysis of this paper, we expect our results to hold, at least 
qualitatively, in that model.}  Their masses are related by 
\beq
M_{T_+} \,=\,M_{T_-}\,\left(1-\frac{m_t^2 f^2}{v^2 M_{T_-}^2}\right)^{-1/2},
\eeq{topmasses}
so that there is just one additional independent parameter in this sector.
We will choose it to be $M_{T_-}$, and assume $M_{T_-}>M$ to avoid a 
charged LTP. The couplings of the heavy photon $B_H$ which will be used in the 
calculations of this paper are summarized in Table~\ref{tab:coup}.

\begin{table}[t!]
\begin{center}
\begin{tabular}{|c|c|} \hline
   $\B^\mu\B^\nu h$ & $-\frac{i}{2}g^{\prime 2} v g^{\mu\nu}$  \\ \hline
   $\B^\mu\B^\nu hh$ &  $-\frac{i}{2}g^{\prime 2} g^{\mu\nu}$ \\ \hline
   $\B^\mu \tilde{Q}_i^a Q_j^b$ & $i \tilde{Y}\,g^\prime 
\gamma^\mu P_L \delta_{ij} \delta^{ab}$ \\ \hline 
   $\B^\mu \tilde{L}_i L_i$ & $i \tilde{Y} \,g^\prime
\gamma^\mu P_L \delta_{ij}$ \\ \hline 
   $\B^\mu T_- t$ & $i \left( \frac{2}{5}\right)\,g^\prime\gamma^\mu
   \sin\alpha\left(\sin\alpha\frac{v}{f} P_L+P_R\right)$ 
   \\ \hline
\end{tabular}
\caption{Interaction vertices involving the heavy photon $B_H$ that appear in 
the calculations of this paper. Here $\alpha=\cos^{-1} (M_{T_-}/M_{T_+})$,
and $\tilde{Y}=1/10$.}
\label{tab:coup}
\end{center}
\end{table}

\section{Relic Density Calculation} 
\label{relic}

\begin{figure}[t]
\begin{center}
\begin{picture}(420,80)(0,-10)
\Text(1,1)[tr]{$B_H$}
\Text(1,61)[br]{$B_H$}
\Photon(1,1)(31,31){4}{4}
\Photon(1,61)(31,31){4}{4}
\Vertex(31,31){2}
\DashLine(31,31)(81,31){5}
\Vertex(81,31){2}
\Photon(81,31)(111,1){4}{4}
\Photon(81,31)(111,61){4}{4}
\Text(56,33)[bc]{$h$}
\Text(111,1)[tl]{$W^+/Z$}
\Text(111,61)[bl]{$W^-/Z$}
\Text(181,1)[tr]{$B_H$}
\Text(181,61)[br]{$B_H$}
\Photon(181,1)(211,31){4}{4}
\Photon(181,61)(211,31){4}{4}
\Vertex(211,31){2}
\DashLine(211,31)(261,31){5}
\Vertex(261,31){2}
\ArrowLine(261,31)(291,61)
\ArrowLine(291,1)(261,31)
\Text(236,33)[bc]{$h$}
\Text(294,1)[tl]{$\bar{t}$}
\Text(294,61)[bl]{$t$}
\Text(331,1)[cr]{$B_H$}
\Text(331,61)[cr]{$B_H$}
\Photon(331,1)(371,1){4}{3}
\Photon(331,61)(371,61){4}{3}
\Vertex(371,1){2}
\Vertex(371,61){2}
\ArrowLine(371,1)(371,61)
\ArrowLine(401,1)(371,1)
\ArrowLine(371,61)(401,61)
\Text(404,1)[cl]{$\bar{q}$}
\Text(404,61)[cl]{$q$}
\Text(368,31)[cr]{$\tilde{Q}$}
\end{picture}
\vskip2mm
\caption{The leading $2\leftrightarrow 2$ processes which maintain the
heavy photon in equilibrium with the rest of the cosmic fluid at high
temperatures.}
\label{fig:annih}
\end{center}
\end{figure}
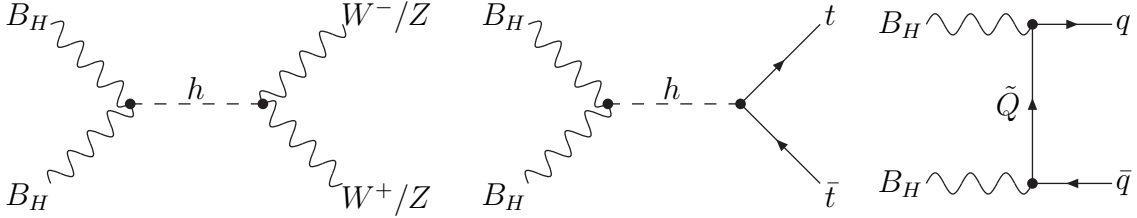

In the early universe, the heavy photons are in equilibrium with the rest of 
the cosmic fluid. In the simplest case of generic (non-degenerate) T-odd 
particle mass spectrum, the equilibrium is maintained via the heavy photon 
pair-annihilation and pair-creation reactions; the leading 
$2\leftrightarrow 2$ processes that contribute 
are shown in Fig.~\ref{fig:annih}. The present relic 
abundance of heavy photons is determined by the behavior of 
pair-annihilation rates in the non-relativistic limit, namely, by the sum of 
the quantities
\beq
a(X)= \lim_{u\to 0} \sigma(B_HB_H\to X)u,   
\eeq{aterm}
over all possible final states $X$. Here, $u$ is the relative
velocity of the annihilating particles. Note that, unlike the bino-like neutralinos typically predicted by the constrained minimal supersymmetric standard model (cMSSM), the $s$-wave annihilation of the heavy photons is unsuppressed: in the language of Ref.~\cite{MIDM}, the heavy photons are ``$s$ annihilators'', analogous to the Kaluza-Klein photons of the ``universal extra dimensions'' (UED) model~\cite{UEDDM,CFM}. 
It is straightforward to compute $a(X)$ using the Feynman rules in
Table~\ref{tab:coup}. We obtain 
\beqa
a (W^+W^-) &=& \frac{2\pi\alpha^2}{3\cw4}
\,\frac{M^2}{(4M^2-m_h^2)^2+m_h^2 \Gamma_h^2}\,
\left(1-\mu_w+\frac{3}{4}\mu_w^2\right)\sqrt{1-\mu_w}\,,\CR
a (ZZ) &=& \frac{\pi\alpha^2}{3\cw4}
\,\frac{M^2}{(4M^2-m_h^2)^2+m_h^2 \Gamma_h^2}\,
\left(1-\mu_z+\frac{3}{4}\mu_z^2\right)\sqrt{1-\mu_z}\,,\CR
\eeqa{wz}
where $\mu_i=m_i^2/M^2$, $\theta_W$ is the SM weak mixing angle, 
and $m_h$ and $\Gamma_h$ are the mass and the width of the SM Higgs boson. 
If $M>m_t$, the LTPs can also annihilate into pairs of top quarks; ignoring 
the contribution from the $t$- and $u$-channel $T_-$ exchanges, we 
obtain\footnote{The $T_-$ exchanges are negligible throughout most of the 
parameter space, but will nevertheless be fully included in the numerical 
calculation of the relic abundance described below.}
\beq
a (t\bar{t})\,=\,\frac{\pi\alpha^2}{\cw4}\,
\frac{M^2}{(4M^2-m_h^2)^2+m_h^2 \Gamma_h^2}\,
\mu_t (1-\mu_t)^{3/2}\,.
\eeq{top}
If $M>m_h$, annihilation into a pair of Higgs bosons is possible, with the 
cross section
\beq
a (hh) = \frac{\pi\alpha^2 M^2}{2\cw4}\,\left[
\frac{\mu_h(1+\mu_h/8)}{(4M^2-m_h^2)^2+m_h^2 \Gamma_h^2}\,+\,\frac{1}{24M^4}
\right]\,\sqrt{1-\mu_h}.
\eeq{higgs}
Finally, LTPs can also annihilate into light SM fermions via $t$-channel 
exchanges of the T-odd fermions; this channel was not included in the analysis 
of Ref.~\cite{HM}. For a fermion $f$ ($f=\ell^\pm, \nu, u, d$) we obtain
\beq
a (f\bar{f})\,=\, \frac{16\pi\alpha^2\tilde{Y}^4N_c^f}{9\cw4}\,
\frac{M^2}{(M^2+\tilde{M}^2)^2}\,,
\eeq{f_i}
where $N_c^f=1$ for leptons and 3 for quarks, and $\tilde{Y}=1/10$ is the $B_H f \tilde{f}$ coupling in units of $g^\prime$. Because of the small value of $\tilde{Y}$, the annihilation into light fermions is strongly suppressed, even for relatively small values of $\tilde{M}$. The WMAP collaboration data~\cite{WMAP}
provides a precise determination of the present dark matter abundance: at 
two-sigma level,
\beq
\Omega_{\rm dm} h^2 = 0.111 \pm 0.018.
\eeq{WMAP}
For 
$s$ annihilators, this translates into a determination of the quantity
$a\equiv \sum_X a(X)$: 
$a=0.8\pm 0.1$ pb. (The precise central value of $a$ depends on the WIMP
mass; however, this dependence is very mild, see Fig.~1 of Ref.~\cite{MIDM}.)
Using this constraint and the above formulas, it is straightforward to map
out the regions of the model parameter space where the heavy photons can 
account for all of the observed dark matter. The results are consistent with the updated analysis of Hubisz and Meade, see Fig.~3 of Ref.~\cite{HM}. For given $m_h$, there are two values of $M$ which result in the correct relic density. 
There is one solution on either side of the Higgs resonance. For WIMP masses in the interesting range, Eq.~\leqn{WIMPmassrange}, these can be approximated by simple analytic expressions:
\beq
m_h \approx 24+2.38 M,~~~{\rm or}~~~m_h \approx -83+1.89 M,
\eeq{solutions}
where $M$ and $m_h$ are in units of GeV. We will refer to these solutions 
as ``low'' and ``high'', respectively. The analytic 
expressions~\leqn{solutions} reproduce the values of $M$ and $m_h$ consistent 
with the WMAP central value of $\Omega_{\rm dm}h^2$ with an error of at most 
a few GeV throughout the interesting parameter range. This accuracy will
be sufficient for the analysis of detection prospects in Sections~\ref{direct}
and~\ref{indirect}.

Throughout the parameter space consistent with the WMAP value of the present
dark matter density, the dominant heavy photon annihilation channels are
$W^+W^-$ and $ZZ$; the $t\bar{t}$ channel contributes at most about 5\% of 
the total annihilation cross section, while the $hh$ final state is always 
kinematically forbidden. Moreover, the ratio of the $W^+W^-$ and $ZZ$
contributions is approximately 2:1, as is evident from Eqs.~\leqn{wz},
so that $a(W^+W^-)\approx 0.53$~pb, $a(ZZ)\approx 0.27$~pb throughout the
parameter space. Since $B_H$ is an $s$-annihilator, the same cross sections
govern the rate of heavy photon annihilation in the galactic halo, which 
in turn determines the fluxes relevant for indirect detection, see
Section~\ref{indirect}. 

If some of the T-odd particles are approximately degenerate in mass with 
the heavy photon, the simple analysis above is no longer applicable, since 
coannihilation reactions between $B_H$ and other states significantly affect 
the relic abundance. In the LHT model, the masses of the T-odd weak gauge 
bosons $W_H$ and the triplet scalar $\phi$ are predicted unambiguously once 
the scale $f$ and the Higgs mass $m_h$ are fixed; these particles are always 
much heavier than the $B_H$ and their effect is negligible. On the other 
hand, the common mass scale of the T-odd leptons and quarks $\tilde{M}$ is a 
free parameter, and for $M\sim \tilde{M}$ the coannihilations between these 
states and the $B_H$ can be important. We have performed a more detailed 
analysis of the $B_H$ relic density, taking this possibility into account. 

In the presence of coannihilations, the abundance calculation requires solving a system of coupled Boltzmann equations. We approached this problem numerically. The interactions of the LHT model were incorporated in the CalcHEP package~\cite{calchep}, which was used to compute the scattering matrix elements for the appropriate processes. The rest of the calculation was performed using the {\tt DM++} package,\footnote{The {\tt DM++} package is currently being prepared for public release. It can be applied to compute the relic abundance of the WIMP for any particle physics model that can be incorporated in CalcHEP. The {\tt DM++} is inspired by the {\tt micrOMEGAs} code~\cite{micrOM}, which was originally designed to compute the relic abundance of neutralinos in the MSSM. The recently developed new version of this code, {\tt micrOMEGAs2.0}~\cite{LH2005}, is also applicable to any CalcHEP model defined by the user. This package is also being prepared for public release.}
recently  recently developed by one of us (AB). The package
first uses the matrix elements to compute the thermal averages $\left< \sigma u\right>$, which determine the reaction rates entering the Boltzmann equation. Then, the freeze-out temperature of the dark matter is determined iteratively, using the Turner-Scherrer approximation~\cite{turner}. Finally, the integral of $\left<\sigma u\right>$ from freeze-out to present day (usually called $J\left( x_F \right)$ in the literature) is evaluated, providing the relic abundance.

\begin{figure}[tb]
\begin{center}
\includegraphics[width=7.5cm]{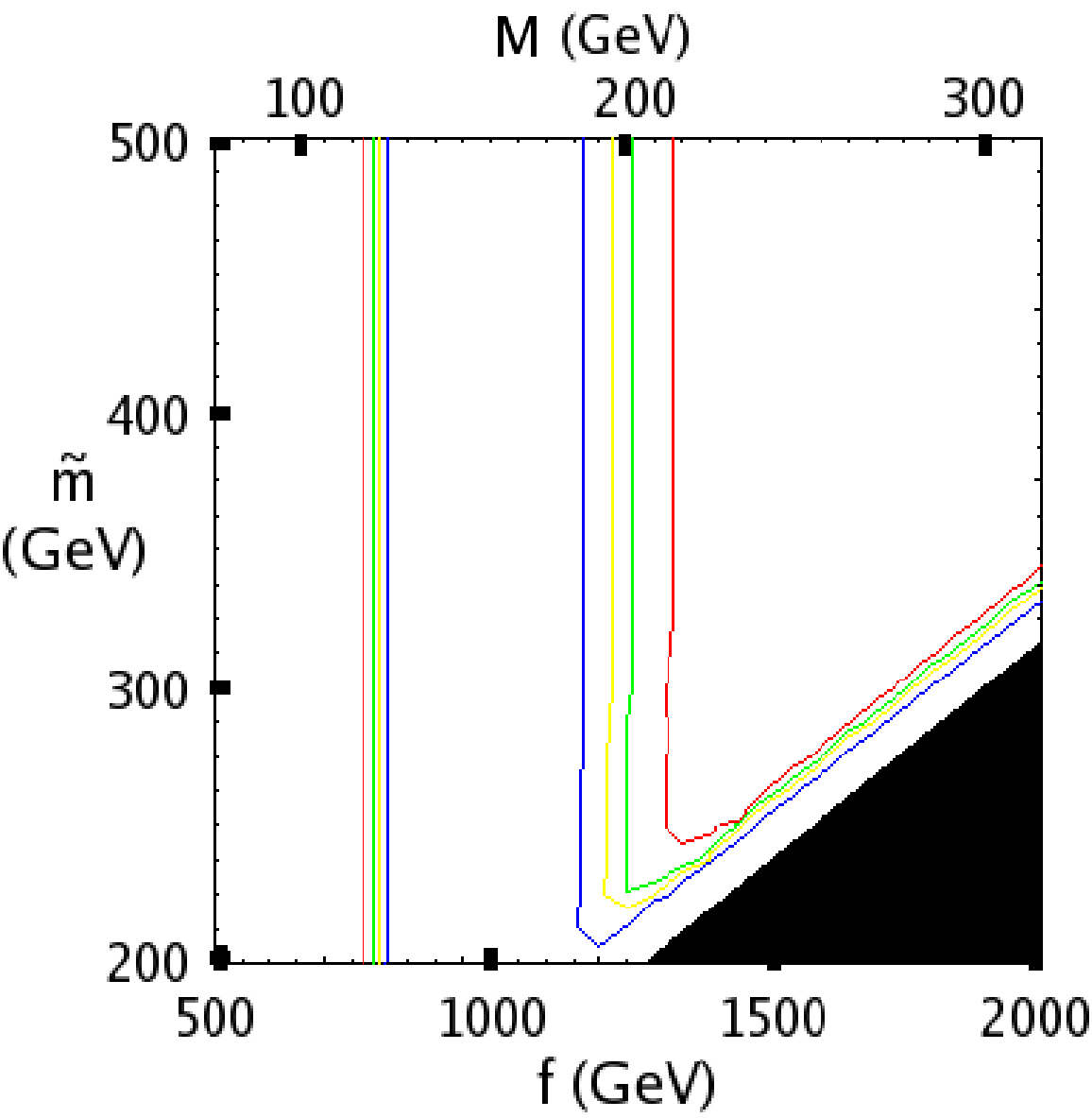}
\includegraphics[width=7.5cm]{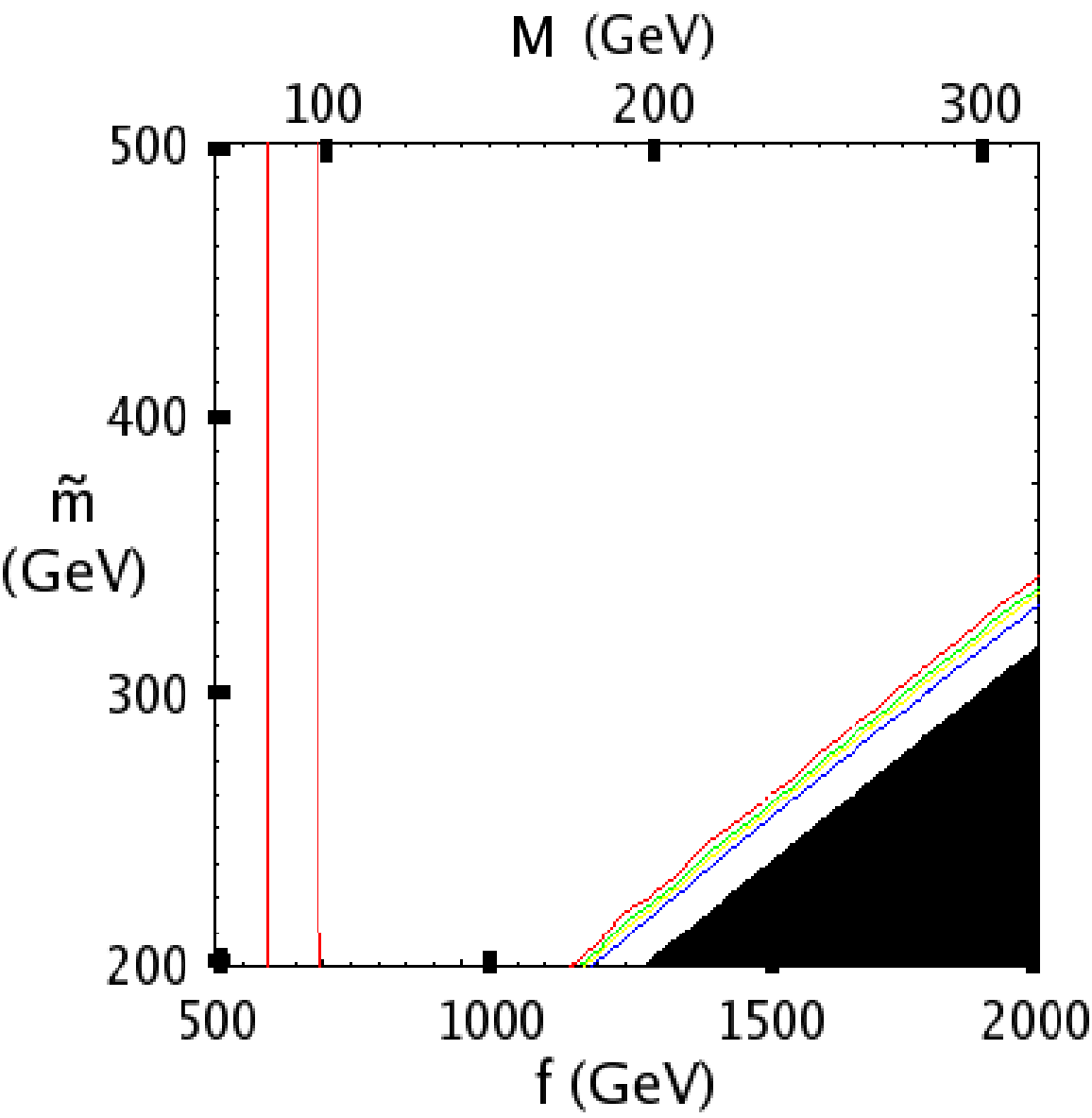}
\vskip2mm
\caption{The contours of constant present abundance of the heavy photon
LTP, $\Omega_{\rm LTP} h^2$, in the $M-\tilde{M}$ plane. The Higgs mass is 
taken to be 300 GeV (left panel) and 120 GeV (right panel). The red and green 
contours correspond to the upper and lower bounds from WMAP, Eq.~\leqn{WMAP},
assuming that the LTP makes up all of dark matter. The yellow 
and blue lines correspond to the LTP contributing 50\% and 70\%,
respectively, of the measured dark matter density. The shaded 
region corresponds to a charged and/or colored LTP.}
\label{fig:coan}
\end{center}
\end{figure}
     
The results of this analysis are illustrated by Figure~\ref{fig:coan}, which shows the contours of constant heavy photon relic density in the $f-\tilde{M}$ (or, equivalently, $M-\tilde{M}$) plane. The typical situation for a heavy
Higgs is shown in the 
left panel ($m_h=300$ GeV). There are two regions in which the heavy photon can account for the observed dark matter:

\begin{itemize}

\item The two vertical {\it pair-annihilation bands}, where the 
coannihilation processes are unimportant. The heavy photon abundance in these
regions is independent of $\tilde{M}$. The bands appear on 
either side of the $s$-channel Higgs resonance dominating the 
pair-annihilation processes, corresponding to the ``high'' and ``low''
solutions of Eq.~\leqn{solutions}. (The bands are analogous to the 
``Higgs funnel'' region in the cMSSM.) 

\item The {\it coannihilation tail}, where the heavy photon abundance 
is predominantly set by coannihilation processes. Since the T-odd fermions are assumed to be degenerate, all of them participate in the coannihilation reactions. The location and shape of this feature are similar to the tau coannihilation tail in cMSSM. 

\end{itemize}

As the Higgs mass is decreased, the pair-annihilation bands appear for lower 
WIMP masses, and for light Higgs (115--150 GeV) the ``low'' band disappears,
since the required values of $f$ are already ruled out by data. The 
``high'' band persists until the Higgs mass is close to the current 
experimental
bound. To illustrate this, consider the right panel of Fig.~\ref{fig:coan},
where $m_h=120$ GeV. The band between the two red lines ($90\lapproxeq M 
\lapproxeq 100$ GeV) is allowed. Note that the behavior of the relic density 
as a function of $M$ within this band is non-trivial: The relic
density first drops with increasing $M$ due to the fact that the threshold 
for the reaction $B_H B_H\to ZZ$ is passed. It then bottoms out at a value
consistent with the measured $\Omega_{\rm dm}h^2$, and begins increasing
as increasing $M$ further takes the center-of-mass energy away from the
Higgs resonance, suppressing annihilation. Clearly, this situation is
quite non-generic, and for somewhat higher $m_h$ the $Z$ threshold 
becomes irrelevant and relic density is a uniformly increasing function of 
$M$ in the ``high'' band. The coannihilation tail is present for low as
well as high values of $m_h$. The tail can be described by a simple analytic 
formula
\beq
\tilde{M}\approx M+20~{\rm GeV},
\eeq{tail}
which is approximately independent of the Higgs mass.

It should be noted that the remaining free parameter of our model, the mass 
of the second T-odd top quark $M_{T_-}$, was fixed to be equal to $f$, so 
that $M_{T_-}\gg M$ and this particle did not have an effect on the $B_H$ 
relic abundance. We expect that a second coannihilation tail appears when 
$M_{T_-}\sim M$; the structure should be very similar to the one found above, 
with slight numerical differences due to smaller multiplicity of the 
coannihilating states.

\section{Direct Detection}
\label{direct}

Direct dark matter detection experiments attempt to observe the recoil energy transfered to a target nucleus in an elastic collision with a WIMP. The null result of the current experiments places an upper bound on the cross section of elastic WIMP-nucleon scattering. In this section, we will discuss the implications of this bound for the LHT dark matter, and prospects for future discovery.

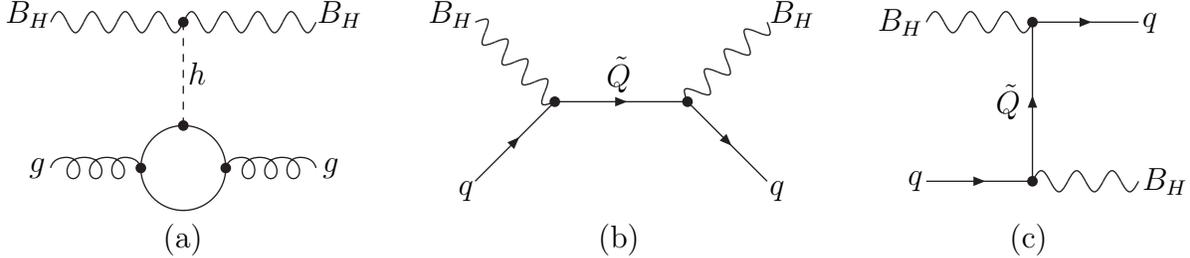
\begin{figure}[t]
\begin{center}
\begin{picture}(420,80)(0,-25)
\Text(1,61)[br]{$B_H$}
\Text(102,61)[bl]{$B_H$}
\Photon(1,61)(51,61){4}{4}
\Photon(51,61)(101,61){4}{4}
\Vertex(51,61){2}
\Gluon(1,6)(35,6){4}{3}
\CArc(51,6)(16,0,360)
\Vertex(35,6){2}
\Vertex(67,6){2}
\Gluon(67,6)(101,6){4}{3}
\DashLine(51,61)(51,22){3}
\Vertex(51,22){2}
\Text(-1,6)[cr]{$g$}
\Text(104,6)[cl]{$g$}
\Text(53,42)[cl]{$h$}
\Text(51,-15)[tc]{(a)}
\Photon(161,61)(191,31){4}{4}
\ArrowLine(161,1)(191,31)
\Vertex(191,31){2}
\ArrowLine(191,31)(241,31)
\Vertex(241,31){2}
\ArrowLine(241,31)(271,1)
\Photon(241,31)(271,61){4}{4}
\Text(273,61)[bl]{$B_H$}
\Text(273,1)[tl]{$q$}
\Text(161,61)[br]{$B_H$}
\Text(161,1)[tr]{$q$}
\Text(216,37)[bc]{$\tilde{Q}$}
\Text(216,-15)[tc]{(b)}
\Text(331,1)[cr]{$q$}
\Text(331,61)[cr]{$B_H$}
\Photon(411,1)(371,1){4}{3}
\Photon(331,61)(371,61){4}{3}
\Vertex(371,1){2}
\Vertex(371,61){2}
\ArrowLine(331,1)(371,1)
\ArrowLine(371,1)(371,61)
\ArrowLine(371,61)(411,61)
\Text(414,1)[cl]{$B_H$}
\Text(414,61)[cl]{$q$}
\Text(368,31)[cr]{$\tilde{Q}$}
\Text(371,-15)[tc]{(c)}
\end{picture}
\vskip2mm
\caption{The leading processes which contribute to the
heavy photon--nucleon elastic scattering cross section relevant for
direct dark matter detection experiments.}
\label{fig:DDproc}
\end{center}
\end{figure}

The elastic scattering of the heavy photon on a nucleus receives contributions from several processes shown in Fig.~\ref{fig:DDproc}. 
Consider first the scattering off gluons, which occurs via the 
Higgs exchange diagram (a). The Higgs-gluon coupling arises predominantly via a top quark loop, and has the form~\cite{Okun}
\beq
{\cal L}_{hgg} \,=\, \frac{\alpha_s}{12\pi v}\,hG^a_{\mu\nu} G^{a\mu\nu}.
\eeq{Hglueglue}
where $v=246$ GeV is the Higgs vev, and $G^a_{\mu\nu}$ is the color field strength. The halo WIMPs are highly nonrelativistic ($\beta\sim 10^{-3}$), and the momentum transfer in the reaction at hand is negligible compared to $m_h$. The WIMP-gluon interaction can then be described by an effective operator
\beq
\frac{\alpha_s \alpha}{6 \cw2} 
\frac{1}{m_h^2} B_{{\rm H}\alpha} B^\alpha_{\rm H}\, G^a_{\mu\nu} G^{a\mu\nu}.
\eeq{SIglue}
In the chiral limit, the matrix element $\left< n|G^2 |n\right>$ can be related to the nucleon mass $\mN$~\cite{Okun}, leading to an effective WIMP-nucleon vertex of the form
\beq
{\cal L}_{\rm eff}\,=\,\frac{e^2}{27\cw2}\frac{\mN}{m_h^2}\,B_{{\rm H}\alpha} B^\alpha_{\rm H}\, \bar{\Psi}_n \Psi_n,
\eeq{NGGN}
where $\Psi_n$ is the nucleon (neutron or proton) field.
It is clear that this interaction only contributes to the spin-independent (SI) part of the WIMP-nucleon scattering cross section. Neglecting other contributions to the SI cross section (which, as we will argue below, are expected to be subdominant), we obtain
\beq
\sigma_{\rm SI}\,=\,\frac{4\pi\alpha^2}{729\cw4}\,\frac{\mN^4}{m_h^4}\,
\frac{1}{(M+m_n)^2},
\eeq{SIxsec}
for both neutrons and protons. Since the scattering off nucleons in a given nucleus is coherent and the matrix elements for neutrons and protons are identical, the SI cross section for scattering off a nucleus of mass $m_N$ is simply obtained from Eq.~\leqn{SIxsec} by a substitution $m_n \to m_N$. 

The interaction of WIMPs with quarks is dominated by the T-odd quark exchange diagrams, see Fig.~\ref{fig:DDproc} (b) and (c). (The Higgs exchange diagrams are suppressed due to small Yukawa couplings of quarks. In fact, it is well known that the Higgs-nucleon interaction is dominated by the Higgs-gluon coupling considered above.) The scattering amplitude is given by 
\beq
- i \frac{e^2\tilde{Y}^2}{\cw2} 
\varepsilon_{\mu}^{\ast}(p_3) \varepsilon_{\nu}(p_1) \,\,
\bar{u} (p_4) \! \left[ 
\frac{\gamma^{\mu} \! \sla{k}_1 \gamma^{\nu}} {k_1^2 - \tilde{M}^2} 
+ \frac{\gamma^{\nu} \! \sla{k}_2 \gamma^{\mu}} {k_2^2 - \tilde{M}^2} 
\right] \! P_L \, u(p_2) \ , 
\eeq{Toddamp}
where $k_1 = p_1 + p_2$, $k_2 = p_2 - p_3$. The $q\tilde{q}B_H$ coupling is flavor-independent, $\tilde{Y}=1/10$, and the expression~\leqn{Toddamp} is valid for every quark species. The amplitude contains two important physical scales: the weak scale, $M\sim \tilde{M} \sim 100$ GeV, and the QCD scale, $\Lambda_{\rm QCD}\sim 100$ MeV, which represents the typical energy and momentum of the quarks bound inside a stationary nucleus and, by a coincidence, the spatial momenta of the halo WIMPs: $|{\bf p}|_{1,3}\sim \beta M \sim \Lambda_{\rm QCD}$. We will work to leading order in the ratio of these two scales. In this approximation,
$k_1\approx -k_2 \approx (M,{\mathbf 0})$, and the heavy photon polarization vectors are purely spatial, $\varepsilon^\mu(p_{1,3})=(0, \varepsilon_{1,3})$. The amplitude takes the form
\beq
\frac{e^2\tilde{Y}^2}{\cw2}\,\frac{M}{M^2-\tilde{M}^2}\,\epsilon_{ijk} \varepsilon_1^i \varepsilon^j_3 \bar{u}_4 \gamma^k (1-\gamma^5) u_2\,,
\eeq{ToddNR}
corresponding to the coupling of the $B_H$ spin with the vector and axial-vector quark currents. The axial current interaction corresponds to the coupling between the WIMP and quark spins, and gives rise to the spin-dependent (SD) part of the WIMP-nucleus scattering cross section. By the Wigner-Eckardt theorem, the quark axial current can be replaced by the nuclear spin operator $s^\mu_N$:
\beq
\left<N| \bar{q}\gamma^\mu \gamma^5 q |N\right> \,=\, 2 s^\mu_N \lambda_q.
\eeq{spin_fla}
For a nucleus of spin $J_N$, the coefficients are given by
\beq
\lambda_q = \Delta q_p \frac{\left<S_p\right>}{J_N} +
\Delta q_n \frac{\left<S_n\right>}{J_N},
\eeq{lambda}
where $\left<S_{p,n}\right>/J_N$ is the fraction of the total nuclear spin 
carried by protons and neutrons, respectively, and the quantities 
$\Delta q_n$ can be extracted from deep inelastic scattering data. We will 
use $\Delta u_p = \Delta d_n = 0.78 \pm 0.02$, $\Delta d_p = \Delta u_n = 
-0.48 \pm 0.02$, $\Delta s_n = \Delta s_p = -0.15\pm 0.02$~\cite{Deltas}. The 
effective WIMP-nucleus spin-spin interaction can then be written as  
\beq
\frac{2e^2\tilde{Y}^2M}{\cw2 (M^2-\tilde{M}^2)}\,\epsilon_{ijk} B_H^i B_H^j 
\bar{\Psi}_N s^k_N \Psi_N\,\sum_{q=u,d,s} \lambda_q,
\eeq{ToddAxi}
yielding the SD cross section
\beq
\sigma_{\rm SD} \,=\, \frac{16\pi\alpha^2\tilde{Y}^4}{3\cw4}  
\frac{m_N^2}{(M+m_N)^2}\,\frac{M^2}{(M^2-\tilde{M}^2)^2}\,J_N (J_N+1) 
\left(\sum_{q=u,d,s} \lambda_q\right)^2.
\eeq{SDxsec}

Now, consider the part of the amplitude~\leqn{ToddNR} involving the quark 
vector current. Since the current is conserved, the contributions of each 
valence quark in a nucleon add coherently, and sea quarks do not contribute. 
The resulting WIMP-nucleon coupling is
\beq
\frac{3e^2\tilde{Y}^2M}{\cw2 (M^2-\tilde{M}^2)}\,\epsilon_{ijk} B_H^i B_H^j 
\bar{\Psi}_n \gamma^k \Psi_n\,.
\eeq{ToddVec}
This interaction is suppressed in the nonrelativistic limit, since 
$\bar{u}_n\gamma^k u_n\sim v_n^k$. In fact, it is of the same order as other 
contributions to the WIMP-quark scattering amplitude, suppressed by WIMP 
velocities or powers of $\Lambda_{\rm QCD}/M$, which were neglected in our 
analysis. Therefore, its effect will be neglected.

\begin{figure}[tb]
\begin{center}
\includegraphics[width=7.3cm]{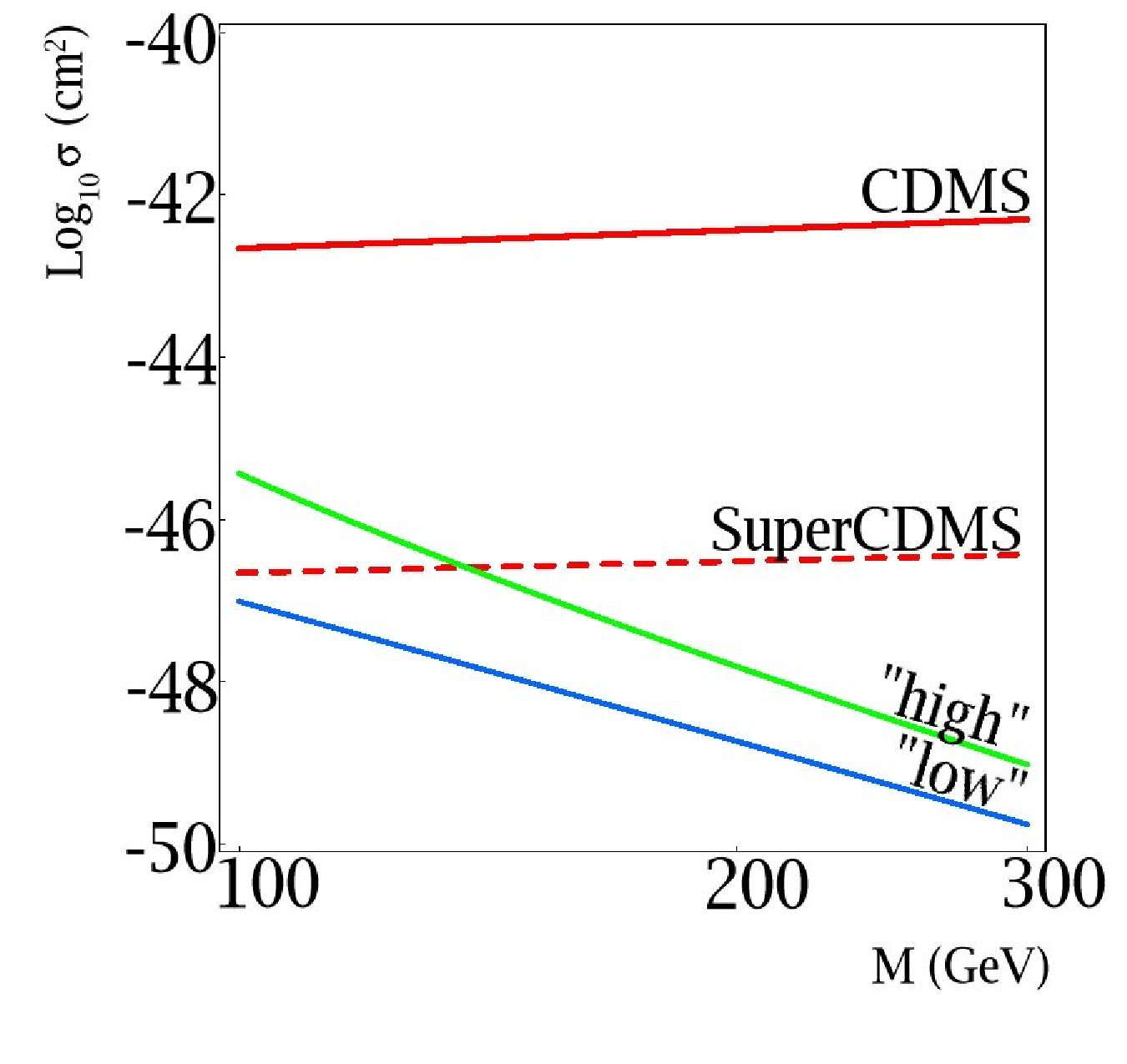}
\hskip.5cm
\includegraphics[width=7.3cm]{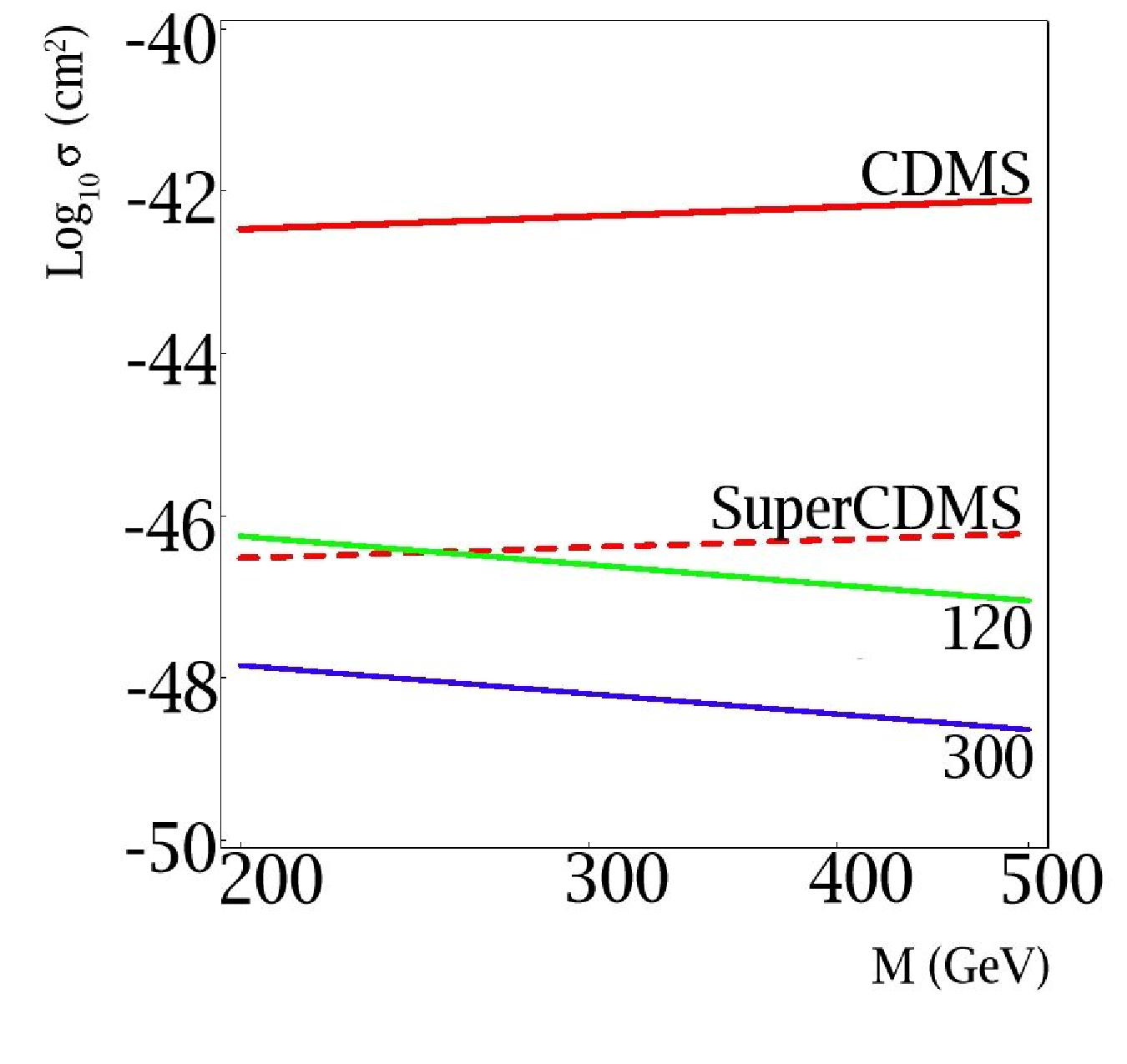}
\vskip2mm
\caption{The spin-independent (SI) WIMP-nucleon elastic scattering cross 
section in the pair-annihilation bands (left panel) and in the coannihilation 
region, for two values of $m_h$, 120 and 300 GeV (right panel). The 
present~\cite{CDMSnow} and projected~\cite{CDMSfut} sensitivities of the CDMS 
experiment are also shown.}
\label{fig:SIdata}
\end{center}
\end{figure}

It should be noted that the SI interaction in Eq.~\leqn{NGGN},  
is parametrically suppressed with respect to the leading SD coupling, Eq.~\leqn{ToddAxi}, by a factor of $m_N/m_h\sim \Lambda_{\rm QCD}/M$, and is formally of the same order as the contributions to the WIMP-quark interaction that were neglected in our analysis. Since the neglected terms contribute to the SI as well as SD interactions, one may question the validity of the SI cross section obtained in Eq.~\leqn{SIxsec}. Note, however, that the WIMP-quark interactions are additionally suppressed by a factor of $\tilde{Y}^2=0.01$, not present in the WIMP-gluon couplings. Thus, while of the same order as~\leqn{NGGN} in terms of power counting, the neglected SI corrections from WIMP-quark interactions are expected to be numerically small. One interesting potential exception occurs in the coannihilation region, where the $\tilde{Y}$ suppression could be compensated by the factor of $\tilde{M}-M \ll M$ in the propagator, and the WIMP-quark interactions could provide a significant correction to Eq.~\leqn{SIxsec}. A detailed analysis of this issue is reserved for future study.

\begin{figure}[tb]
\begin{center}
\includegraphics[width=7.3cm]{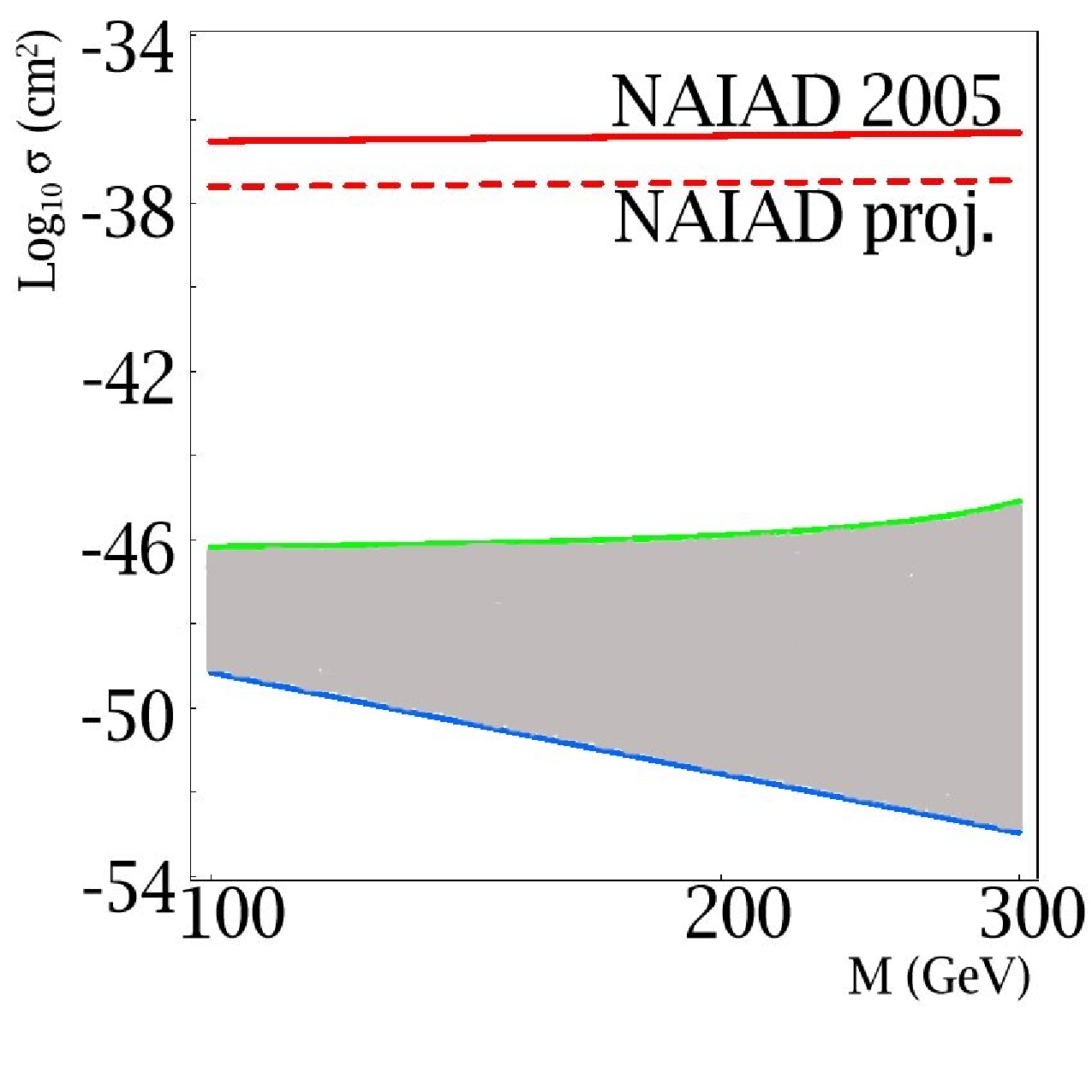}
\hskip.5cm
\includegraphics[width=7.3cm]{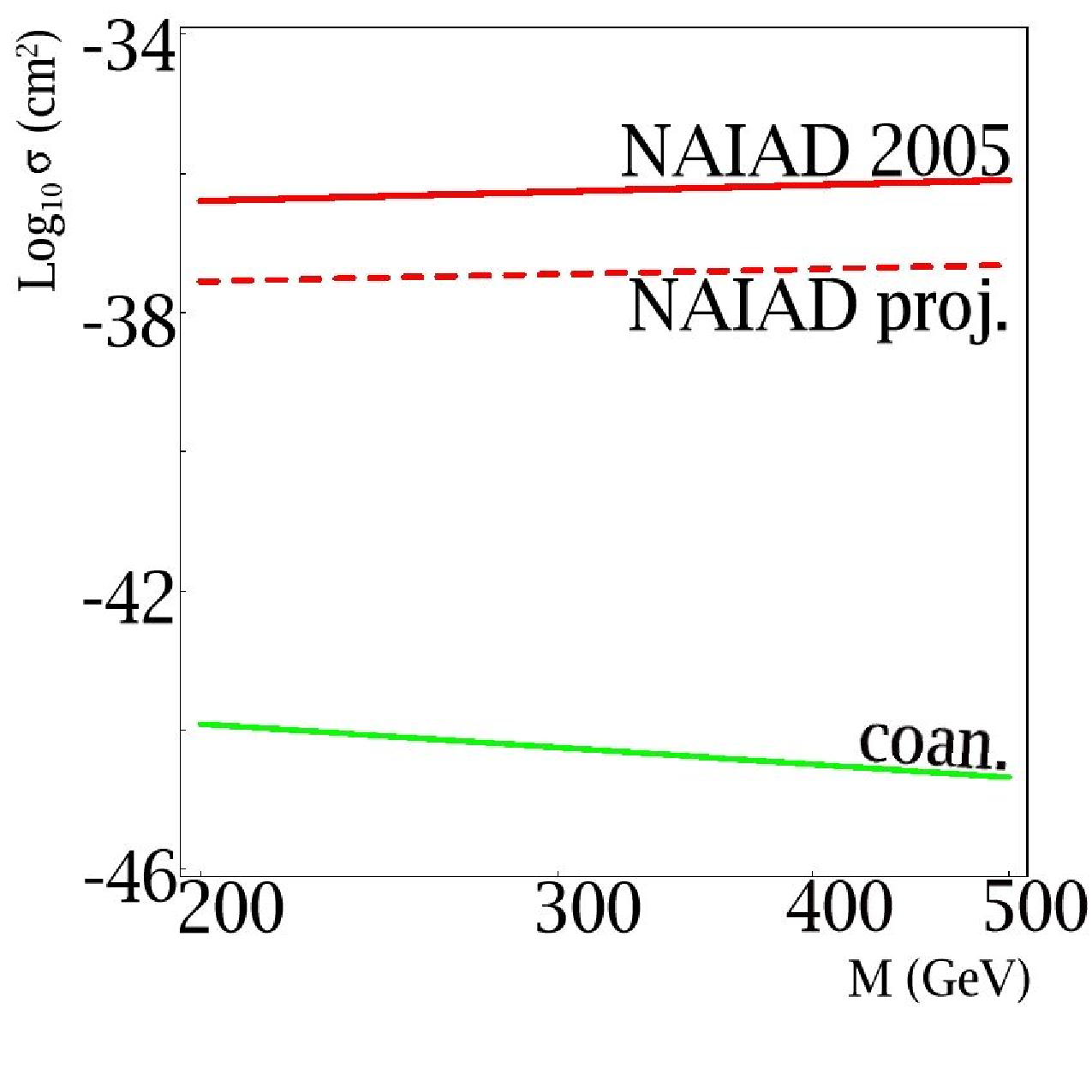}
\vskip2mm
\caption{The spin-dependent (SD) WIMP-proton elastic scattering cross 
section in the pair-annihilation bands (left panel) and in the coannihilation 
region (right panel). In the pair-annihilation bands, the scale $\tilde{M}$
is allowed to vary between 350 GeV and the upper bound given in 
Eq.~\leqn{upper}. The present~\cite{Naiadnow} and projected~\cite{Naiadfut} 
sensitivities of the NAIAD experiment are also shown.}
\label{fig:SDdata}
\end{center}
\end{figure}

The SI elastic WIMP-nucleon scattering cross sections expected in the LHT 
models are plotted in Fig.~\ref{fig:SIdata}, along with the current bound
from the CDMS collaboration~\cite{CDMSnow} (solid red lines) and the
projected future sensitivity of SuperCDMS, stage C~\cite{CDMSfut} 
(dashed red lines). We assume that the heavy photons account for all of the 
observed dark matter, and the two panels correspond to the two regions of 
parameter space which satisfy this constraint. The left panel shows the 
cross section expected in the pair-annihilation bands, with the two lines 
corresponding to the high and low solutions in Eq.~\leqn{solutions}.
The right panel shows the cross section expected
in the coannihilation tail for two values of the Higgs mass, 120 GeV and 
300 GeV. The two lines can be thought of as the upper 
and lower bounds on the expected cross section.\footnote{Note, however, 
that in the LHT model a heavy Higgs, $m_h>300$ GeV, may be consistent with 
precision electroweak data in certain 
regions of parameter space where its contibution to the $T$ parameter is 
partially cancelled by new physics contributions~\cite{HMNP}. A heavier 
Higgs corresponds to smaller SI cross section.} While the predicted cross
sections are two-three orders of magnitude below the present sensitivity,
the expected improvements of the CDMS experiments will 
allow it to begin probing the interesting regions of the model parameter 
space in both pair-annihilation and coannihilation regions.

Fig.~\ref{fig:SDdata} shows the spin-dependent cross sections predicted by the
LHT model, along with the current bound from the NAIAD 
experiment~\cite{Naiadnow} and its projected sensitivity~\cite{Naiadfut}.
In the pair-annihilation bands, the scale $\tilde{M}$
is allowed to vary between 350 GeV and the upper bound given in 
Eq.~\leqn{upper}. (Recall that for a given value of $M$, the scale $f$ is 
fixed unambiguously by Eq.~\leqn{mass}.) Unfortunately, the predicted SD 
cross sections are several orders of magnitude below the NAIAD sensitivity.

\section{Indirect Detection via Anomalous Gamma Rays}
\label{indirect}

As discussed in Section~\ref{relic}, WIMP annihilation processes have to 
occur with approximately weak-scale cross sections to ensure that the relic 
abundance of WIMPs is consistent with observations. Since the heavy photons 
of the LHT model are $s$-annihilators, their annihilation rates are 
approximately velocity-independent in the nonrelativistic regime. This 
implies that the WIMPs collected, for example, in galactic halos, have a 
substantial probability to pair-annihilate, resulting in anomalous high-energy 
cosmic rays which could be distinguished from astrophysical 
backgrounds. In particular, high-energy gamma rays (photons) and positrons 
are considered to be the most promising experimental signatures. The gamma
ray signal is particularly interesting because the gamma rays in the relevant 
energy range travel over galactic scales with no scattering, so that if
the signal is observed, information about the WIMP (e.g. its mass)
could be extracted from the spectrum. In this section, we will compute the 
gamma ray fluxes predicted by the LHT model, and evaluate their 
observability.\footnote{Positron fluxes from the heavy photon dark matter
annihilation in the LHT model were recently considered in 
Ref.~\cite{positrons}.}

There are three principal mechanisms by which hard photons can be produced 
in WIMP annihilation:

\begin{itemize}
\item  Monochromatic photons produced via direct annihilation into a two body 
final state ($\gamma\gamma, h\gamma$ or $Z\gamma$);

\item  Photons radiated in the process of hadronization and fragmentation of 
strongly interacting particles produced either directly in WIMP annihilation 
(e.g. $B_HB_H\to q\bar{q}$) or in hadronic decays of the primary annihilation 
products (e.g. $B_HB_H\to ZZ$ followed by $Z\to q\bar{q}$);

\item  Photons produced via radiation from a final state charged particle.

\end{itemize}

Let us consider each of these mechanisms in turn in the LHT model.

\begin{figure}[t]
\begin{center}
\begin{picture}(360,80)(0,-10)
\Text(1,1)[tr]{$B_H$}
\Text(1,61)[br]{$B_H$}
\Photon(1,1)(31,31){4}{4}
\Photon(1,61)(31,31){4}{4}
\Vertex(31,31){2}
\DashLine(31,31)(81,31){5}
\Vertex(81,31){2}
\Photon(81,31)(111,1){4}{4}
\Photon(81,31)(111,61){4}{4}
\Photon(111,1)(111,61){4}{4}
\Vertex(111,1){2}
\Vertex(111,61){2}
\Photon(111,1)(151,1){4}{4}
\Photon(111,61)(151,61){4}{4}
\Text(56,33)[bc]{$h$}
\Text(153,1)[cl]{$\gamma$}
\Text(153,61)[cl]{$\gamma$}
\Text(115,31)[cl]{$W$}
\Text(201,1)[tr]{$B_H$}
\Text(201,61)[br]{$B_H$}
\Photon(201,1)(231,31){4}{4}
\Photon(201,61)(231,31){4}{4}
\Vertex(231,31){2}
\DashLine(231,31)(281,31){5}
\Vertex(281,31){2}
\ArrowLine(281,31)(311,61)
\ArrowLine(311,1)(281,31)
\ArrowLine(311,61)(311,1)
\Photon(311,1)(351,1){4}{4}
\Photon(311,61)(351,61){4}{4}
\Vertex(311,1){2}
\Vertex(311,61){2}
\Text(256,33)[bc]{$h$}
\Text(353,1)[cl]{$\gamma$}
\Text(353,61)[cl]{$\gamma$}
\Text(315,31)[cl]{$t$}
\end{picture}
\vskip2mm
\caption{The diagrams which dominate the monochromatic photon 
pair-production in the $B_H$ annihilation in the galactic halo.}
\label{fig:gg}
\end{center}
\end{figure}
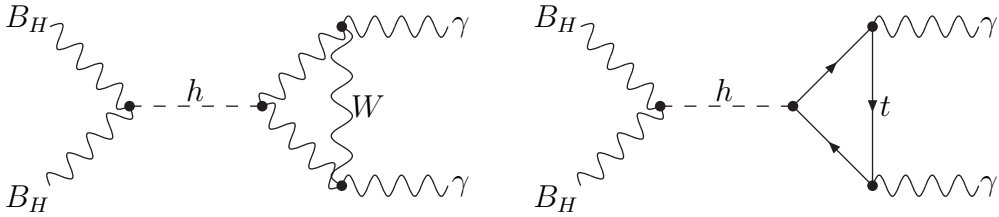

WIMPs being electrically neutral, production of monochromatic photons can 
only occur at loop level. In this paper, we will concentrate on the 
$\gamma\gamma$ final state, postponing the analysis of the $Z\gamma$ and 
$h\gamma$ channels for future work. (The photons produced in these reactions
are separated in energy from the $\gamma\gamma$ photons due to non-zero 
masses of the $Z$ and the Higgs.) The $\B\B\to\gamma\gamma$ process 
is dominated by the one-loop diagrams inducing the effective $h\gamma\gamma$ 
vertex, see 
Fig.~\ref{fig:gg}.\footnote{A complete calculation would also include the 
contribution of the box diagrams with T-odd and T-even quarks running in the
loop, analogous to the quark/squark boxes entering in the case of MSSM
neutralino annihilation~\cite{2gammaMSSM}. In the LHT case, this contribution
is expected to be subdominant since the matrix element contains a factor of 
$\tilde{Y}^2=0.01$.} 
The corresponding cross section can be easily evaluated using the well-known 
formulas for the Higgs boson partial widths:
\beq
\sigma_{\gamma\gamma} u \equiv \sigma\left(B_HB_H \to \gamma\gamma\right)u = 
\frac{g^{\prime 4}v^2}{72 M^4}\,
\frac{s^2-4s M^2+12 M^4}{(s-m_h^2)^2+m_h^2\Gamma_h^2}\, 
\frac{\hat{\Gamma} \left( h \to V_1 V_2\right)}{\sqrt{s}},
\eeq{higgsphoton}
where $u$ is the relative velocity of the annihilating WIMPs, and $s\approx 
4M^2$ in the non-relativistic regime relevant for the galactic WIMP
annihilation. The 
hat on $\Gamma$ indicates that the substitution $m_h\to\sqrt{s}$ should be 
performed in the standard expressions for on-shell Higgs 
decays~\cite{Russians,Hunters}, and the loops of new particles present in the 
LHT model should be included. We obtain
\beq
\hat{\Gamma}(h\to\gamma\gamma)\,=\,\frac{\alpha^2 g^2}{1024 \pi^3} 
\frac{s^{3/2}}{m_W^2}\,\Bigl| {\cal A}_1 + {\cal A}_{1/2} + 
{\cal A}_0 \Bigr|^2,
\eeq{sum}
where ${\cal A}_s$ denotes the contribution from loops of particles of spin $s$. These contributions are given by
\beqa
{\cal A}_1 &=& \sum_i c_i Q_i^2 \frac{\tau_W}{\tau_i} F_1(\tau_i); \CR
{\cal A}_{1/2} &=& \sum_i \frac{\sqrt{2} Q_i^2 y_i }{g} 
\frac{\tau_W^{1/2}}{\tau_i^{1/2}} F_{1/2}(\tau_i); \CR
{\cal A}_0 &=& \sum_i \frac{2 Q_i^2 \lambda_i}{g^2} \frac{\tau_W}{\tau_i} 
F_0(\tau_i),
\eeqa{spinamps}
where the sums run over all the charged particles of a given spin, and 
implicitly include summations over colors and other quantum numbers where 
necessary. The particles in the sums have masses $m_i$ and electric charges 
(in units of the electron charge) $Q_i$; their trilinear couplings to the 
Higgs boson are given by $\lambda_i v$, $y_i/\sqrt{2}$, and 
$c_i g M_W \eta^{\mu\nu}$, for particles of spin 0, 1/2, and 1, respectively. 
(With these normalization choices, $c_i=1$ for the SM $W^\pm$, and $y_i$'s 
are the usual Yukawas for the SM fermions.) We have also defined 
$\tau_i=4m_i^2/s$. The functions $F_s(\tau)$ are given by
\beqa
F_1(\tau) &=& 2\plus 3\tau\plus 3\tau\left( 2\minus\tau\right) f(\tau), \CR
F_{1/2}(\tau) &=& \minus 2\tau \left( 1\plus\left( 1\minus\tau\right) f(\tau )\right), \CR
F_0(\tau) &=& \tau\left( 1\minus\tau f(\tau )\right), 
\eeqa{Fs}
where
\beqa
f(\tau ) &=& \left[ \sin^{-1}\left(\sqrt{\frac{1}{\tau}}\right)\right]^2~~{\rm if}~\tau > 1, \CR
& & 
-\frac{1}{4}\left[\log\left(\frac{1\plus\sqrt{1\minus\tau}}{1\minus\sqrt{1\minus\tau}}\right)\minus i\pi\right]^2~{\rm if}~\tau < 1.
\eeqa{ftau}
Using these expressions, we find that the 
contributions of the T-odd states are subdominant compared to the SM
loops. The contributions of the T-odd fermion loops and the T-even 
heavy top loop are suppressed because their coupling to the Higgs is of 
order $v^2/f^2$. The contributions of charged T-odd heavy gauge bosons and 
scalars are suppressed due to their large masses, of order $f$. The 
deviation of the effective $h\gamma\gamma$ coupling from its Standard Model
value due to these states is of order a few per cent.\footnote{The deviations 
of the $h\to \gamma\gamma$ and $gg\to h$ vertices from the SM in the LHT model 
were recently analyzed in detail in Ref.~\cite{hgg}.} Given the much larger
astrophysical uncertanties inherent in the anomalous photon flux predictions,  
we will ignore these effects in our analysis.
 
The monochromatic flux due to the
$\gamma\gamma$ final state, observed by a telescope with a line of sight 
parametrized by $\Psi=(\theta,\varphi)$ and a field of view $\Delta\Omega$ 
can be written as~\cite{Buckley}
\beq
\Phi \,=\,  \left(1.1 \times 10^{-9} {\rm s}^{-1}{\rm cm}^{-2}\right)\,\left( 
\frac{\sigma_{\gamma\gamma} u}{1{\rm~pb}} \right)\,
\left(\frac{100~{\rm GeV}}{M}\right)^2 \bar{J}(\Psi,\Delta\Omega)
\Delta\Omega.
\eeq{lineflux}
The function $\bar{J}$ contains the dependence of the flux on the halo dark 
matter density distribution:
\beq
\bar{J}(\Psi, \Delta\Omega ) \equiv \frac{1}{8.5~{\rm kpc}} \left( 
\frac{1}{0.3 {\rm~GeV/cm}^3} \right)^2 \frac{1}{\Delta\Omega} 
\int_{\Delta\Omega} d\Omega \int_\Psi \rho^2 dl.
\eeq{jbar}
where $l$ is the distance from the observer along the line of sight. Many 
models of the galactic halo predict a sharp peak in the dark matter density 
in the neighborhood of the galactic center, making the line of sight towards 
the center the preferred one for WIMP searches.\footnote{Note, however, that 
a powerful point-like source of ultra high energy gamma rays has been 
recently detected in the galactic center region~\cite{point}. The energy 
spectrum of this source, smooth and extending out to at least a few TeV, 
makes its interpretation in terms of WIMP annihilation unlikely. Detection 
of the potential gamma flux from WIMP annihilation in the same spatial region 
is clearly made more difficult by the presence of the source.}
However, the features of the 
predicted peak are highly model-dependent, resulting in a large uncertainty 
in the predicted $\bar{J}$. For example, at $\Delta\Omega=10^{-3}$~sr, 
characteristic of ground-based Atmospheric Cerenkov Telescopes (ACTs), typical 
values of $\bar{J}$ range from $10^3$ for the NFW profile~\cite{NFW} to 
about $10^5$ for the profile of Moore et.al.~\cite{Moore}, and can be further 
enhanced by a factor of up to $10^2$ due to the effects of adiabatic 
compression~\cite{abc}.


\begin{figure}[tb]
\begin{center}
\includegraphics[width=7.5cm]{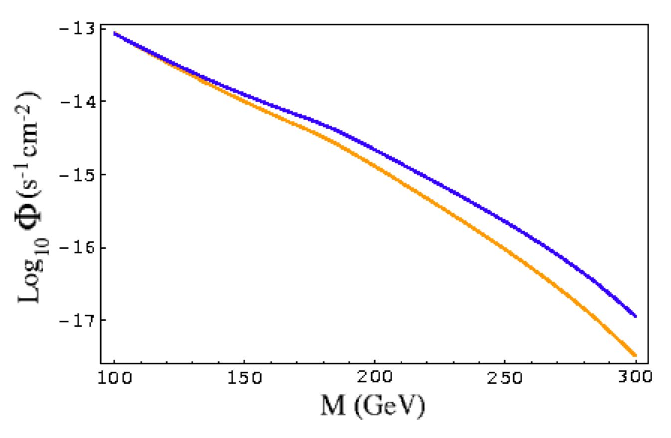}
\hskip.5cm
\includegraphics[width=7.0cm]{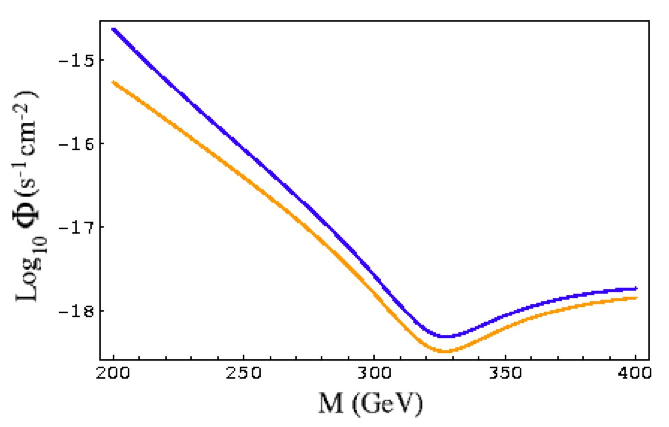}
\vskip2mm
\caption{The flux of the monochromatic photons from the reaction 
$B_HB_H\rightarrow \gamma\gamma$ in the pair-annihilation bands (left panel) 
and the coannihilation region (right panel). On the left panel, the blue/upper 
and the orange/lower lines correspond to the high and low solutions in 
Eq.~\leqn{solutions}, respectively. On the right panel, the blue/upper line 
corresponds to $m_h=300$ GeV, the orange/lower line to $m_h=120$ GeV. The 
plots assume $\bar{J}(\Psi, \Delta\Omega)\Delta\Omega=1$; all fluxes scale
linearly with this parameter.}
\label{fig:xsecgg}
\end{center}
\end{figure}

The monochromatic photon fluxes (assuming $\bar{J} \Delta\Omega=1$)
predicted by the LHT model in the parameter regions where the heavy photon 
accounts for all of the observed dark matter, are shown in 
Fig.~\ref{fig:xsecgg}. The left panel 
corresponds to the pair-annihilation bands, and the right panel to the 
coannihilation region. Searches for gamma rays from WIMP annihilation have 
to be able to 
distinguish them from the astrophysical background. In the case of the 
monochromatic photons, the signal is concentrated in a single bin (the
energy uncertainty of the telescopes is about 10\%, much larger than
the intrinsic line width), and the background can be effectively measured
in the neighbouring bins and subtracted. In the relevant energy range,
the flux sensitivity for ground-based Atmospheric Cherenkov Telescopes (ACTs) 
such as VERITAS~\cite{VERITAS} and HESS~\cite{HESS} is 
estimated to be around $(1-5)\times 10^{-12}$ cm$^{-2}$sec$^{-1}$, whereas 
the sensitivity of the
upcoming space-based telescope GLAST is limited by statistics at 
$10^{-10}$~cm$^{-2}$sec$^{-1}$, assuming that 10 events are required to
claim discovery~\cite{GLASTsens}. It is clear that the monochromatic
flux predicted by 
the LHT model is beyond the reach of GLAST, but could be observed at the 
ACTs if the dark matter distribution in the halo exhibits a 
substantial spike or strong clumping, $\bar{J}\gapproxeq 10^5$ at 
$\Delta\Omega\approx 10^{-3}$.

\begin{figure}[t]
\begin{center}
\includegraphics[width=10cm]{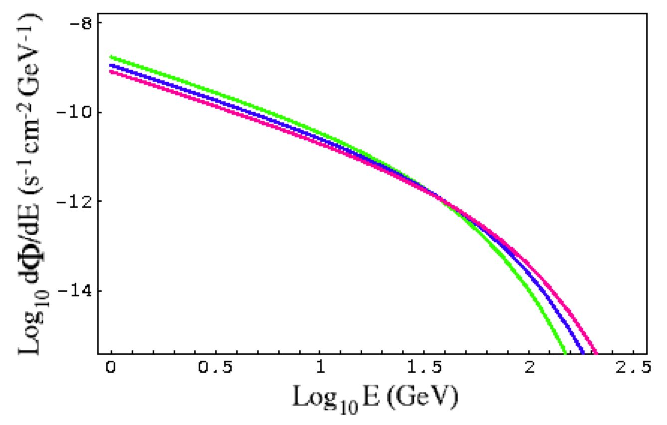}
\vskip2mm
\caption{The fragmentation photon flux for $M=150, 200, 250$ GeV 
(green, blue and red lines, respectively), in the pair-annihilation bands. 
The plot assumes $\bar{J}(\Psi, \Delta\Omega)\Delta\Omega=1$; all fluxes 
scale linearly with this parameter.}
\label{fig:fragflux}
\end{center}
\end{figure}

Let us now consider the component of the photon flux due to 
hadronization and fragmentation of quarks produced in WIMP annihilation.
As discussed in Section~\ref{relic},
the heavy photons predominantly annihilate into $W$ and $Z$ pairs; each of the
vector bosons can in turn decay into a quark pair. The resulting photon 
spectra depend only on the initial energies of the $W$'s and $Z$'s, and not 
on the details of the WIMP annihilation process. The spectra have 
been studied using {\tt PYTHIA} (in the MSSM context), and a simple analytic 
fit has been presented in Ref.~\cite{Buckley}:
\beq
\frac{dN_\gamma}{dx} \approx \frac{0.73}{x^{1.5}}e^{-7.8x}.
\eeq{blatantlystolen}
where $x=E_\gamma/M$. This approximation is valid for both $W^+W^-$ and 
$ZZ$ final states. In the pair-annihilation bands, the differential flux is 
then given by
\beq
\frac{d\Phi}{dE} \,=\, \left( 3.3\times 10^{-12} {\rm s}^{-1}{\rm cm}^{-2}
{\rm GeV}^{-1}\right)\,x^{-1.5}e^{-7.8x}\, 
\left(\frac{100~{\rm GeV}}{M}\right)^3 \bar{J}(\Psi,\Delta\Omega)
\Delta\Omega,
\eeq{Fragflux}
where we used the relic density constraint, $a(W^+W^-)+a(ZZ)\approx 0.8$
pb. The flux in the coannihilation region is much smaller.

The fluxes predicted by Eq.~\leqn{Fragflux} for several values of $M$ are 
plotted in Fig.~\ref{fig:fragflux}. The GLAST telescope is
statistics-limited at energies above about 2 GeV, and would observe tens
of events in this energy range for the heavy photon mass in the preferred
range, assuming $\bar{J}(\Delta\Omega)\Delta\Omega=1$. (The flux scales
linearly with this parameter combination.) One should keep in mind,
however, that while the prospects for observing this signal are good, 
ruling out its interpretation in terms of conventional astrophysics could be
challenging given the smooth, featureless nature of the fragmentation 
spectrum. Detailed studies of the angular distribution of these photons, 
in particular outside the galactic disk, will be needed. 

The ACTs have a higher energy threshold, typically about 50 GeV, and 
suffer from an irreducible background from electron-induced showers, about 
$10^{-12}-10^{-10}$~cm$^{-2}$s$^{-1}$GeV$^{-1}$ in the relevant energy range 
($50\ldots 200$ GeV) for $\Delta\Omega\sim 10^{-3}$. Using the 
extraploation of Ref.~\cite{Buckley} to estimate the background, we find that
the typical signal/background ratio expected at the ACTs, assuming 
$\Delta\Omega\sim 10^{-3}$ and $\bar{J}(\Delta\Omega)\Delta\Omega=1$, is
only about $10^{-3}$. An observation of the fragmentation flux at the 
ACTs appears quite challenging, unless dark matter is strongly clustered
at the galactic center or clumped.

The third and final component of the gamma-ray flux from WIMP annihilation 
is the final state radiation (FSR) photons. The FSR flux generally 
provides a robust signature of WIMP annihilation: it exists whenever 
the WIMPs have a sizable annihilation cross section into {\it any} charged 
states. The FSR photons have a continuous spectrum, in analogy to the
quark fragmentation photons consdered above. In fact, at low
energies, the fragmentation flux dominantes over the FSR
component (unless WIMPs annihilate into purely leptonic states). At 
energies close to the WIMP mass, however, the fragmentation flux drops 
sharply, and the FSR component typically dominates~\cite{endpoint}. This is   
particularly interesting because the FSR spectrum typically possesses a sharp 
edge feature, abruptly dropping to zero at the maximal photon energy 
allowed by kinematics. The edge feature could help the experiments to 
discern this flux on top of the (a priori highly uncertain) 
astrophysical background, and provide a measurement of the WIMP 
mass~\cite{endpoint}. In the LHT model, the dominant charged two-body 
annihilation channel is $W^+W^-$, and correspondingly the reaction 
$B_HB_H \to W^\plus W^\minus\gamma$ provides the most important component of 
the FSR photon flux.  The differential cross section for this process is 
given by
\beq
\frac{d\sigma}{dx}\left( B_HB_H\to W^\plus W^\minus\gamma\right) =  
\sigma\left( B_HB_H \to W^\plus W^\minus\right) \,{\cal F}(x; \mu_w),
\eeq{WWgamma}
where $x=2E_\gamma/\sqrt{s}\approx E_\gamma/M$, $\mu_w= (m_W/M)^2$ and
\beqa
{\cal F}(x;\mu) &=&\frac{\alpha}{\pi}  \frac{1}{\sqrt{1-\mu}}\,\frac{1}{x}\,
\times \Bigl[ (2x-2+\mu)\log\frac{2(1-x)-\mu-2\sqrt{(1-x)(1-x-\mu)}}{\mu} 
\CR & & 
+ 2\left(\frac{8x^2}{4-4\mu+3\mu^2}-1\right)\sqrt{(1-x)(1-x-\mu)}
\Bigr]\,,
\eeqa{Factor}
for $0\leq x\leq 1-\mu$ and 0 for $1-\mu\leq x\leq 1$.
In the limit of large heavy photon mass, $s\gg M_W^2$, this expression 
reduces to
\beq
{\cal F}(x) =\frac{2\alpha}{\pi} \frac{1-x}{x} \,\Bigl[
\log\frac{s(1-x)}{m_W^2}\,+\,2x^2-1+{\cal O}(\mu)\Bigr].
\eeq{Factor@mu0}
The leading (logarithmically enhanced) term agrees with the result obtained 
in Ref.~\cite{endpoint} using the Goldstone boson equivalence theorem. 
Note that the form of the photon emission 
factor ${\cal F}$, even in the large-$s$ limit, depends on the theory being 
considered and on the initial state. For example, the 
$\tilde{\chi}_1^0\tilde{\chi}_1^0\to W^+W^-\gamma$ cross section in the MSSM, 
computed in Ref.~\cite{WWsusy}, has a different leading logarithm behavior; 
this is related to the fact that the $W$ bosons effectively become massless 
in this limit, inducing new infrared singularities. Thus, even though we
chose to write the cross section~\leqn{WWgamma} in a ``factorized'' form, 
there is no true factorization in the $WW\gamma$ final state, in contrast 
to the $f\bar{f}\gamma$ final states~\cite{endpoint}.

\begin{figure}[t]
\begin{center}
\includegraphics[width=10cm]{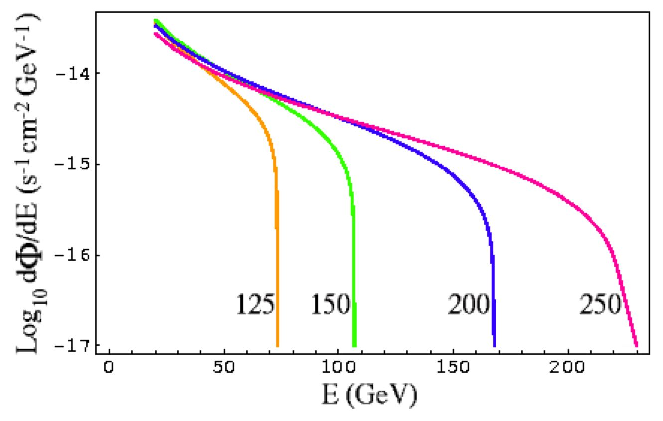}
\vskip2mm
\caption{The FSR photon flux for $M=125, 150, 200, 250$ GeV (left to right),
in the pair-annihilation bands. (The fluxes for ``high'' and ``low'' solutions
are essentially identical.) The plot assumes $\bar{J}(\Psi, \Delta\Omega)
\Delta\Omega=1$; all fluxes scale linearly with this parameter.}
\label{fig:FSRflux}
\end{center}
\end{figure}

The flux of the FSR photons is given by
\beq
\frac{d\Phi}{dE} \,=\, \left( 5.6\times 10^{-12} {\rm s}^{-1}{\rm cm}^{-2}
{\rm GeV}^{-1}\right)\,\left( 
\frac{a(W^+W^-)}{1{\rm~pb}} \right)\,{\cal F}(x; \mu_w) \, 
\left(\frac{100~{\rm GeV}}{M}\right)^3 \bar{J}(\Psi,\Delta\Omega)
\Delta\Omega,
\eeq{FSRflux}
where $a(W^+W^-)$ is given in Eq.~\leqn{wz}. Flux predictions for several
representative values of $M$ are shown in Fig.~\ref{fig:FSRflux}. 
In each case, the flux drops abruptly at the maximal photon energy, 
\beq
E^\gamma_{\rm max} \,=\, \frac{M^2-m_W^2}{M}.
\eeq{egmax}
If this edge feature is observed, it would provide a robust 
signature of heavy photon annihilation, as well as a measurement of its mass.
Note that the FSR and fragmentation components of the flux are comparable 
near the edge, so that the fractional drop in the ``signal'' flux at the edge 
is significant. 
Just as for the monochromatic and fragmentation photons, the sensitivity to 
the FSR flux at ACTs is 
limited by the background from electron-induced showers. 
Assuming a 10\% uncertainty on the total flux measurement, the drop in the 
total flux associated with the edge feature of the FSR spectrum can be 
observed if $\bar{J}\gapproxeq 10^5-10^6$. The sensitivity at GLAST is 
limited by statistics, and an observation of the FSR edge requires even 
higher values of $\bar{J}$. 


To summarize, we found that the best prospects for a discovery of 
anomalous gamma rays due to heavy photon annihilation in the Milky Way are
offered by the GLAST telescope, which should be able to observe tens of
fragmentation photons in the multi-GeV energy range. The fluxes of
monochromatic and FSR photons, whose spectra would provide clear signatures
for galactic WIMP annihilation (a bump and an edge, respectively),
are significantly smaller. The prospects for their detection depend on the 
assumed halo profile; an observation by the ACTs such as VERITAS and HESS 
is possible if the dark matter density has a sharp peak at the galactic 
center or is strongly clumped, $\bar{J}\gapproxeq 10^5-10^6$. 

\section{Conclusions}

Little Higgs models provide an interesting alternative scenario for physics 
at the TeV scale, with a simple and attractive mechanism of radiative
electroweak symmetry breaking. Many realistic models implementing the Little 
Higgs mechanism have been proposed; however, generically these models are 
ruled out by precision electroweak data, unless the scale $f$ is in a
few-TeV range which reintroduces fine-tuning. Little Higgs models with T 
parity avoid this difficulty. In this paper, we focused on the Littlest Higgs 
model with T parity (LHT), one of the simplest models in this class. T
parity makes the lightest of the T-odd particles, the LTP, stable, enabling 
it to have a substantial abundance in today's universe in spite of its 
weak-scale mass. In the LHT model, the LTP is typically the heavy photon 
$\B$, which can play the role of WIMP dark matter. We have computed the 
relic abundance of this particle, including coannihilation effects, and
mapped out the regions of the parameter space where it has the correct 
relic abundance to account for all, or a substantial part, of the observed dark
matter. These regions can be divided into the pair-annihilation bands, 
where the abundance is set by the $\B$ pair annihilation via $s$-channel
Higgs resonance, and the coannihilation tail, where coannihilations of
$\B$ with T-odd quarks $\tilde{Q}$ and leptons $\tilde{L}$ play the
dominant role. 

In the second part of the paper, we evaluated the prospects for observing
the heavy photon dark matter of the LHT model using direct and indirect
detection techniques. Direct detection is quite difficult, due to the fact
that the heavy photon predominantly couples to the Standard Model states 
via the Higgs boson whose interactions with nucleons are weak. The elastic 
cross 
section of the $\B$ scattering on a nucleus in the region of parameter space 
consistent with the relic density constraint was found to be several orders of 
magnitude below the current sensitivity of direct detection searches
such as CDMS. For indirect detection, we concentrated on the anomalous 
high-energy gamma ray signature. The predicted gamma ray flux depends 
sensitively on the distribution of dark matter in the halo. The best 
discovery prospect is offered by the GLAST telescope, which can observe 
the photons arising from the fragmentation of the $W/Z$ bosons produced in
the heavy photon annihilation. If dark matter distribution in the halo is 
favorable (in particular if it ehxibits a sharp spike near the galactic 
center, or is highly clumpy on short scales), ground-based telescopes 
such as VERITAS and HESS may also be able to observe a gamma ray signal. 
In this case, it might also be possible to observe the monochromatic and the
FSR components of the photon flux, whose spectra exhibit well-defined
features (a line and an edge, respectively) and would provide a
smoking-gun evidence for the WIMP-related nature of the signal.  


{\it Acknowledgments ---} We are grateful to Jay Hubisz and Patrick Meade 
for useful discussions. MP, AN and AS are supported by the
NSF grant PHY-0355005. 

MP is grateful to Prateek Agrawal and Can Kilic for pointing out a typo in Eq.~\leqn{top} in the original version of the paper. 
None of the results beyond this formula were affected.


\begin{thebibliography}{99}

\bibitem{DMreview}
For a recent review and a collection of references, see
G.~Bertone, D.~Hooper and J.~Silk,
Phys.\ Rept.\  {\bf 405}, 279 (2005)
[arXiv:hep-ph/0404175].

\bibitem{littlest}   
 N.~Arkani-Hamed, A.~G.~Cohen, E.~Katz and A.~E.~Nelson,
  JHEP {\bf 0207}, 034 (2002)
  [arXiv:hep-ph/0206021]. 
 
\bibitem{LHrev1}  
 M.~Schmaltz and D.~Tucker-Smith,
  arXiv:hep-ph/0502182

\bibitem{LHrev2} 
 M.~Perelstein,
  arXiv:hep-ph/0512128.

\bibitem{pheno}
  G.~Burdman, M.~Perelstein and A.~Pierce,
  Phys.\ Rev.\ Lett.\  {\bf 90}, 241802 (2003)
  [Erratum-ibid.\  {\bf 92}, 049903 (2004)]
  [arXiv:hep-ph/0212228];
  T.~Han, H.~E.~Logan, B.~McElrath and L.~T.~Wang,
  Phys.\ Rev.\ D {\bf 67}, 095004 (2003)
  [arXiv:hep-ph/0301040];
M.~Perelstein, M.~E.~Peskin and A.~Pierce,
  Phys.\ Rev.\ D {\bf 69}, 075002 (2004)
  [arXiv:hep-ph/0310039].

\bibitem{PEW}
  C.~Csaki, J.~Hubisz, G.~D.~Kribs, P.~Meade and J.~Terning,
  Phys.\ Rev.\ D {\bf 67}, 115002 (2003)
  [arXiv:hep-ph/0211124];
 J.~L.~Hewett, F.~J.~Petriello and T.~G.~Rizzo,
  JHEP {\bf 0310}, 062 (2003)
  [arXiv:hep-ph/0211218].

\bibitem{LHT}
  H.~C.~Cheng and I.~Low,
  JHEP {\bf 0309}, 051 (2003)
  [arXiv:hep-ph/0308199];
  JHEP {\bf 0408}, 061 (2004)
  [arXiv:hep-ph/0405243].

\bibitem{Low}
 I.~Low,
  JHEP {\bf 0410}, 067 (2004)
  [arXiv:hep-ph/0409025].

\bibitem{HMNP}
  J.~Hubisz, P.~Meade, A.~Noble and M.~Perelstein,
  JHEP {\bf 0601}, 135 (2006)
  [arXiv:hep-ph/0506042].  

\bibitem{Simplest}
  D.~E.~Kaplan and M.~Schmaltz,
  JHEP {\bf 0310}, 039 (2003)
  [arXiv:hep-ph/0302049];
  M.~Schmaltz,
  JHEP {\bf 0408}, 056 (2004)
  [arXiv:hep-ph/0407143].

\bibitem{Martin}
  A.~Martin,
  arXiv:hep-ph/0602206.

\bibitem{AJ}
  A.~Birkedal-Hansen and J.~G.~Wacker,
  Phys.\ Rev.\ D {\bf 69}, 065022 (2004)
  [arXiv:hep-ph/0306161].

\bibitem{HM}
  J.~Hubisz and P.~Meade, arXiv:hep-ph/0411264, v3.
  See also Phys.\ Rev.\ D {\bf 71}, 035016 (2005); note however that the dark 
matter relic density plot has not been updated in the journal version.

\bibitem{LHTflavor}
  J.~Hubisz, S.~J.~Lee and G.~Paz,
  arXiv:hep-ph/0512169.

\bibitem{noplus}
  H.~C.~Cheng, I.~Low and L.~T.~Wang,
  arXiv:hep-ph/0510225.

\bibitem{MIDM}
  A.~Birkedal, K.~Matchev and M.~Perelstein,
  Phys.\ Rev.\ D {\bf 70}, 077701 (2004)
  [arXiv:hep-ph/0403004].
  
\bibitem{UEDDM}
  G.~Servant and T.~M.~P.~Tait,
  Nucl.\ Phys.\ B {\bf 650}, 391 (2003)
  [arXiv:hep-ph/0206071].

\bibitem{CFM}
 H.~C.~Cheng, J.~L.~Feng and K.~T.~Matchev,
  Phys.\ Rev.\ Lett.\  {\bf 89}, 211301 (2002)
  [arXiv:hep-ph/0207125].

\bibitem{WMAP}
  D.~N.~Spergel {\it et al.}  [WMAP Collaboration],
  Astrophys.\ J.\ Suppl.\  {\bf 148}, 175 (2003)
  [arXiv:astro-ph/0302209].

 \bibitem{calchep} 
  A.~Pukhov,
  arXiv:hep-ph/0412191.
   
\bibitem{micrOM} 
  G.~Belanger, F.~Boudjema, A.~Pukhov and A.~Semenov,
  arXiv:hep-ph/0405253.
  
\bibitem{LH2005}
  B.~C.~Allanach {\it et al.},
  arXiv:hep-ph/0602198, pp.~146--149.

\bibitem{turner}
  R.~J.~Scherrer and M.~S.~Turner,
  Phys.\ Rev.\ D {\bf 33}, 1585 (1986)
  [Erratum-ibid.\ D {\bf 34}, 3263 (1986)].
 
\bibitem{Okun}
  L.~B.~Okun, Leptons and Quarks (1982), pp.  228-231.

\bibitem{Deltas}
  G.~K.~Mallot,
in {\it Proc. of the 19th Intl. Symp. on Photon and Lepton Interactions at High Energy LP99 } ed. J.A. Jaros and M.E. Peskin,
  Int.\ J.\ Mod.\ Phys.\ A {\bf 15S1}, 521 (2000)
  [eConf {\bf C990809}, 521 (2000)]
  [arXiv:hep-ex/9912040].

\bibitem{CDMSnow}
  D.~S.~Akerib {\it et al.}  [CDMS Collaboration],
  Phys.\ Rev.\ Lett.\  {\bf 96}, 011302 (2006)
  [arXiv:astro-ph/0509259].

\bibitem{CDMSfut}
SuperCDMS (Projected) Phase C [from the Dark Matter Plotter web site, 
{\tt http://dmtools.berkeley.edu/limitplots/}].

\bibitem{Naiadnow}
  G.~J.~Alner {\it et al.}  [UK Dark Matter Collaboration],
  Phys.\ Lett.\ B {\bf 616}, 17 (2005)
  [arXiv:hep-ex/0504031].

\bibitem{Naiadfut}
  N.~J.~C.~Spooner {\it et al.},
  Phys.\ Lett.\ B {\bf 473}, 330 (2000).

\bibitem{positrons}
  M.~Asano, S.~Matsumoto, N.~Okada and Y.~Okada,
  arXiv:hep-ph/0602157.

\bibitem{2gammaMSSM}
  L.~Bergstrom and P.~Ullio,
  Nucl.\ Phys.\ B {\bf 504}, 27 (1997)
  [arXiv:hep-ph/9706232];
  Z.~Bern, P.~Gondolo and M.~Perelstein,
  Phys.\ Lett.\ B {\bf 411}, 86 (1997)
  [arXiv:hep-ph/9706538].

\bibitem{Russians}
 M.~A.~Shifman, A.~I.~Vainshtein, M.~B.~Voloshin and V.~I.~Zakharov,
  Sov.\ J.\ Nucl.\ Phys.\  {\bf 30}, 711 (1979)
  [Yad.\ Fiz.\  {\bf 30}, 1368 (1979)].

\bibitem{Hunters}
  J.~F.~Gunion, H.~E.~Haber, G.~L.~Kane and S.~Dawson,
{\it The Higgs Hunter's Guide,} Perseus, Cambridge, MA, 1990;
see also  
  arXiv:hep-ph/9302272.

\bibitem{hgg}
  C.~R.~Chen, K.~Tobe and C.~P.~Yuan,
  arXiv:hep-ph/0602211.

\bibitem{Buckley}
  L.~Bergstrom, P.~Ullio and J.~H.~Buckley,
  Astropart.\ Phys.\  {\bf 9}, 137 (1998)
  [arXiv:astro-ph/9712318].

\bibitem{point}
  K.~Kosack {\it et al.}  [The VERITAS Collaboration],
  Astrophys.\ J.\  {\bf 608}, L97 (2004)
  [arXiv:astro-ph/0403422];
 K.~Tsuchiya {\it et al.}  [CANGAROO-II Collaboration],
  Astrophys.\ J.\  {\bf 606}, L115 (2004)
  [arXiv:astro-ph/0403592];
  F.~Aharonian {\it et al.}  [The HESS Collaboration],
  Astron.\ Astrophys.\  {\bf 425}, L13 (2004)
  [arXiv:astro-ph/0408145].

\bibitem{NFW}
  J.~F.~Navarro, C.~S.~Frenk and S.~D.~M.~White,
  Astrophys.\ J.\  {\bf 490}, 493 (1997).

\bibitem{Moore}
  B.~Moore, F.~Governato, T.~Quinn, J.~Stadel and G.~Lake,
  Astrophys.\ J.\  {\bf 499}, L5 (1998)
  [arXiv:astro-ph/9709051];
B.~Moore, T.~Quinn, F.~Governato, J.~Stadel and G.~Lake,
  Mon.\ Not.\ Roy.\ Astron.\ Soc.\  {\bf 310}, 1147 (1999)
  [arXiv:astro-ph/9903164].

\bibitem{abc}
  G.~R.~Blumenthal, S.~M.~Faber, R.~Flores and J.~R.~Primack,
  Astrophys.\ J.\  {\bf 301}, 27 (1986);
  F.~Prada, A.~Klypin, J.~Flix, M.~Martinez and E.~Simonneau,
  arXiv:astro-ph/0401512;
  O.~Y.~Gnedin, A.~V.~Kravtsov, A.~A.~Klypin and D.~Nagai,
  Astrophys.\ J.\  {\bf 616}, 16 (2004)
  [arXiv:astro-ph/0406247].

\bibitem{VERITAS}
  T.~C.~Weekes {\it et al.},
  Astropart.\ Phys.\  {\bf 17}, 221 (2002)
  [arXiv:astro-ph/0108478].

\bibitem{HESS}
  J.~A.~Hinton  [The HESS Collaboration],
  New Astron.\ Rev.\  {\bf 48}, 331 (2004)
  [arXiv:astro-ph/0403052].

\bibitem{GLASTsens}
  A.~Morselli, A.~Lionetto, A.~Cesarini, F.~Fucito and P.~Ullio  [GLAST
                  Collaboration],
  Nucl.\ Phys.\ Proc.\ Suppl.\  {\bf 113}, 213 (2002)
  [arXiv:astro-ph/0211327].

\bibitem{endpoint}
  A.~Birkedal, K.~T.~Matchev, M.~Perelstein and A.~Spray,
  arXiv:hep-ph/0507194.

\bibitem{WWsusy}
  L.~Bergstrom, T.~Bringmann, M.~Eriksson and M.~Gustafsson,
  Phys.\ Rev.\ Lett.\  {\bf 95}, 241301 (2005)
  [arXiv:hep-ph/0507229].


\end{thebibliography}
\end{document}